\documentclass[sigconf]{acmart}
\usepackage{nicefrac}
\usepackage{todonotes}
\usepackage{enumitem}

\makeatletter
\newcommand{\vast}{\bBigg@{4}}
\newcommand{\Vast}{\bBigg@{5}}
\makeatother

\renewcommand\footnotetextcopyrightpermission[1]{} 

\AtBeginDocument{%
  \providecommand\BibTeX{{%
    \normalfont B\kern-0.5em{\scshape i\kern-0.25em b}\kern-0.8em\TeX}}}

\setcopyright{None}
\copyrightyear{2020}
\acmYear{2020}
\acmDOI{10.1145/XXXXXXX}

\acmConference[AdKDD '20]{AdKDD 2020 Workshop
in conjunction with
The 26th ACM SIGKDD Conference on Knowledge
Discovery and Data Mining}{Aug 2020}{Online} 
\acmBooktitle{AdKDD '20: AdKDD 2020 Workshop,
  Aug 24, 2020, Online}
\acmPrice{15.00}
\acmISBN{978-1-4503-XXXX-X/18/06}

\begin{document}

\title[An Evaluation Framework for Personalization Strategy Experiment Designs]{An Evaluation Framework for Personalization Strategy Experiment Designs}

\author{C. H. Bryan Liu}
\email{bryan.liu12@imperial.ac.uk}
\orcid{0000-0002-6516-2364}
\affiliation{%
  \institution{Imperial College London \& ASOS.com, UK}
}

\author{Emma J. McCoy}
\email{}
\orcid{}
\affiliation{%
  \institution{Imperial College London, UK}
}

\renewcommand{\shortauthors}{Liu and McCoy}

\begin{abstract}
Online Controlled Experiments (OCEs) are the gold standard in evaluating the effectiveness of changes to websites. An important type of OCE evaluates different personalization strategies, which present challenges in low test power and lack of full control in group assignment. We argue that getting the right experiment setup---the allocation of users to treatment/analysis groups---should take precedence of post-hoc variance reduction techniques in order to enable the scaling of the number of experiments. We present an evaluation framework that, along with a few simple rule of thumbs, allow experimenters to quickly compare which experiment setup will lead to the highest probability of detecting a treatment effect under their particular circumstance.
\end{abstract}




\maketitle

\section{Introduction}
\label{sec:introduction}

The use of Online Controlled Experiments (OCEs, e.g. A/B tests) has become popular in measuring the impact of digital products and services, and guiding business decisions on the Web~\cite{liu20whatisthevalue}. Major companies report running thousands of OCEs on any given day~\cite{kohavi13online,xu15frominfrastructure,hohnhold15focusing} and many startups exist purely to manage OCEs~\cite{johari17peeking,browne17whatworks}.

A large number of OCEs address simple variations on elements of the user experience based on random splits, e.g. showing a different colored button to users based on a user ID hash bucket. Here, we are interested in experiments that compare \emph{personalization strategies}, complex sets of targeted user interactions that are common in e-commerce and digital marketing, and measure the change to a performance metric of interest (\emph{metric} hereafter). Examples of personalization strategies include the scheduling, budgeting and ordering of marketing activities directed at a user based on their purchase history. 

Experiments for personalization strategies face two unique challenges. Firstly, strategies are often only applicable to a small fraction of the user base, and thus many simple experiment designs suffer from either a lack of sample size / statistical power, or diluted metric movement by including irrelevant samples~\cite{deng15diluted}.
Secondly, as users are not randomly assigned \emph{a priori}, but must qualify to be treated with a strategy via their actions or attributes, groups of users subjected to different strategies cannot be assumed to be statistically equivalent and hence are not directly comparable.

While there are a number of variance reduction techniques (including stratification and control variates \cite{deng13improving,poyarkov16boosted}) that partially address the challenges, the strata and control variates involved can vary dramatically from one personalization strategy experiment to another, requiring many \emph{ad hoc} adjustments. As a result, such techniques may not scale well when organizations design and run hundreds or thousands of experiments at any given time.

We argue that personalization strategy experiments should focus on the assignment of users from the strategies they qualified for to the treatment/analysis groups. We call this mapping process an \emph{experiment setup}. Identifying the best experiment setup increases the chance to detect any treatment effect. An experimentation framework can also reuse and switch between different setups quickly with little custom input, ensuring the operation can scale. More importantly, the process does not hinder the subsequent application of variance reduction techniques, meaning that we can still apply the techniques \textit{post hoc} if required.

To date, many experiment setups exist to compare personalization strategies. An increasingly popular approach is to compare the strategies using multiple control groups --- Quantcast calls it a dual control~\cite{quantcast}, and Facebook calls it a multi-cell lift study~\cite{liu2018designing}. In the two-strategy case, this involves running two experiments on two random partitions of the user base in parallel, with each experiment further splitting the respective partition into treatment/control and measuring the \emph{incrementality} (the change in a metric as compared to the case where nothing is done) of each strategy. The incrementality of the strategies are then compared against each other. 

Despite the setup above gaining traction in display advertising, there is a lack of literature on whether it is a better setup---one that has a higher sensitivity and/or apparent effect size than other setups. While~\cite{liu2018designing} noted that multi-cell lift studies require a large number of users, they did not discuss how the number compares to other setups.\footnote{A single-cell lift study is often used to measure the incrementality of a single personalization strategy, and hence is not a representative comparison.} The ability to identify and adopt a better experiment setup can reduce the required sample size, and hence enable more cost-effective experimentation.

We address the gap in the literature by introducing an evaluation framework that compares experiment setups given two personalization strategies.
The framework is designed to be flexible so that it is able to deal with a wide range of baselines and changes in user responses presented by any pairs of strategies (\emph{situations} hereafter).
We also recognize the need to quickly compare common setups, and provide some simple rule of thumbs on situations where a setup will be better than another. 
In particular, we outline the conditions where employing a multi-cell setup, as well as metric dilution, is desirable.

To summarize, our contributions are: (i) We develop a flexible evaluation framework for personalization strategy experiments, where one can compare two experiment setups given the situation presented by two competing strategies (Section~\ref{sec:eval_framework}); (ii) We provide simple rule of thumbs to enable experimenters who do not require the full flexibility of the framework to quickly compare common setups (Section~\ref{sec:comparison}); and (iii) We make our results useful to practitioners by making the code used in the paper (Section~\ref{sec:experiments}) publicly available.\footnote{The code and supplementary document are available on: \url{https://github.com/liuchbryan/experiment\_design\_evaluation}.}

\section{Evaluation Framework}
\label{sec:eval_framework}

We first present our evaluation framework for personalization strategy experiments. The experiments compare two personalization strategies, which we refer to as strategy 1 and strategy 2. 
Often one of them is the existing strategy, and the other is a new strategy we intend to test and learn from. 
In this section we introduce (i) how users qualifying themselves into strategies creates non-statistically equivalent groups, (ii) how experimenters usually assign the users, and (iii) when we would consider an assignment to be better.

\subsection{User grouping}
\label{sec:user_grouping}

\begin{figure}
\begin{center}
    \includegraphics[width=0.35\textwidth, trim = 0 0 0 0]{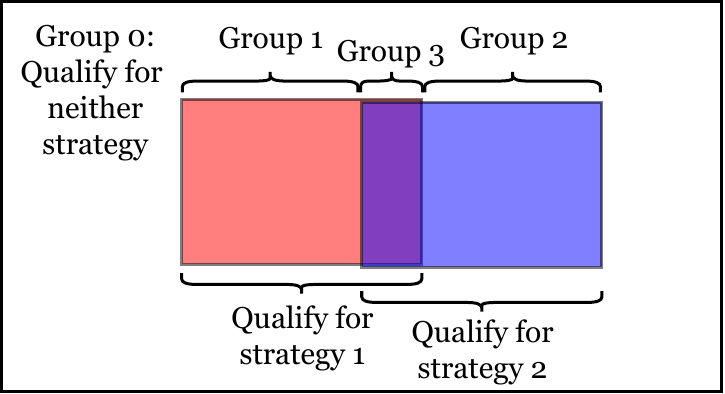}
\end{center}
\caption{Venn diagram of the user groups in our evaluation framework. The outer, left inner (red), and right inner (blue) boxes represent the entire user base, those who qualify for strategy 1, and those who qualify for strategy 2 respectively.}
\label{fig:ME_groups}
\end{figure}

As users qualify themselves into the two strategies, four disjoint groups emerge: those who qualify for neither strategy, those who qualify only for strategy 1, those who qualify only for strategy 2, and those who qualify for both strategies. We denote these groups (user) groups 0, 1, 2, and 3 respectively (see Figure~\ref{fig:ME_groups}). It is perhaps obvious that we cannot assume those in different user groups are statistically equivalent and compare them directly.

We assume groups 0, 1, 2, 3 have $n_0$, $n_1$, $n_2$, and $n_3$ users respectively. We also assume responses from users (which we aggregate, often by taking the mean, to obtain our metric) are distributed differently between groups, and within the same group, between the scenario where the group is subjected to the treatment associated to the corresponding strategy and where nothing is done (baseline). We list all group-scenario combinations in Table~\ref{tab:group_id}, and denote the mean and variance of the responses $(\mu_G, \sigma^2_G)$ for a combination~$G$.\footnote{For example, the responses for group~1 without any interventions has mean and variance ($\mu_{C1}$, $\sigma^2_{C1}$), and that for group~2 with the treatment prescribed under strategy~2 has mean and variance ($\mu_{I2}$, $\sigma^2_{I2}$).}

\begin{table*}
\begin{center}
\begin{tabular}{|c|c|c|c|c|}
\hline
    &  Group 0 & Group 1  & Group 2  & Group 3 \\\hline
 Baseline (\textbf{C}ontrol) & $C0$ & $C1$ & $C2$ & $C3$ \\\hline
 Under treatment (\textbf{I}ntervention) & $/$ & $I1$ & $I2$ & $\begin{array}{ll} \textrm{Under strategy 1:} \quad I\phi \\ \textrm{Under strategy 2:}  \quad I\psi \end{array} $\\
\hline
\end{tabular}
\end{center}
\vspace*{4pt}
\caption{All group-scenario combinations in our evaluation framework for personalization strategy experiments. The columns represent the groups described in Figure~\ref{fig:ME_groups}. The baseline represents the scenario where nothing is done. We assume those who qualify for both strategies (Group 3) can only receive treatment(s) associated to either strategy.}
\label{tab:group_id}
\end{table*}

\subsection{Experiment setups}
\label{sec:experiment_setup}
\begin{figure}
\begin{center}
    \includegraphics[width=0.48\textwidth, trim = 0 0 0 0]{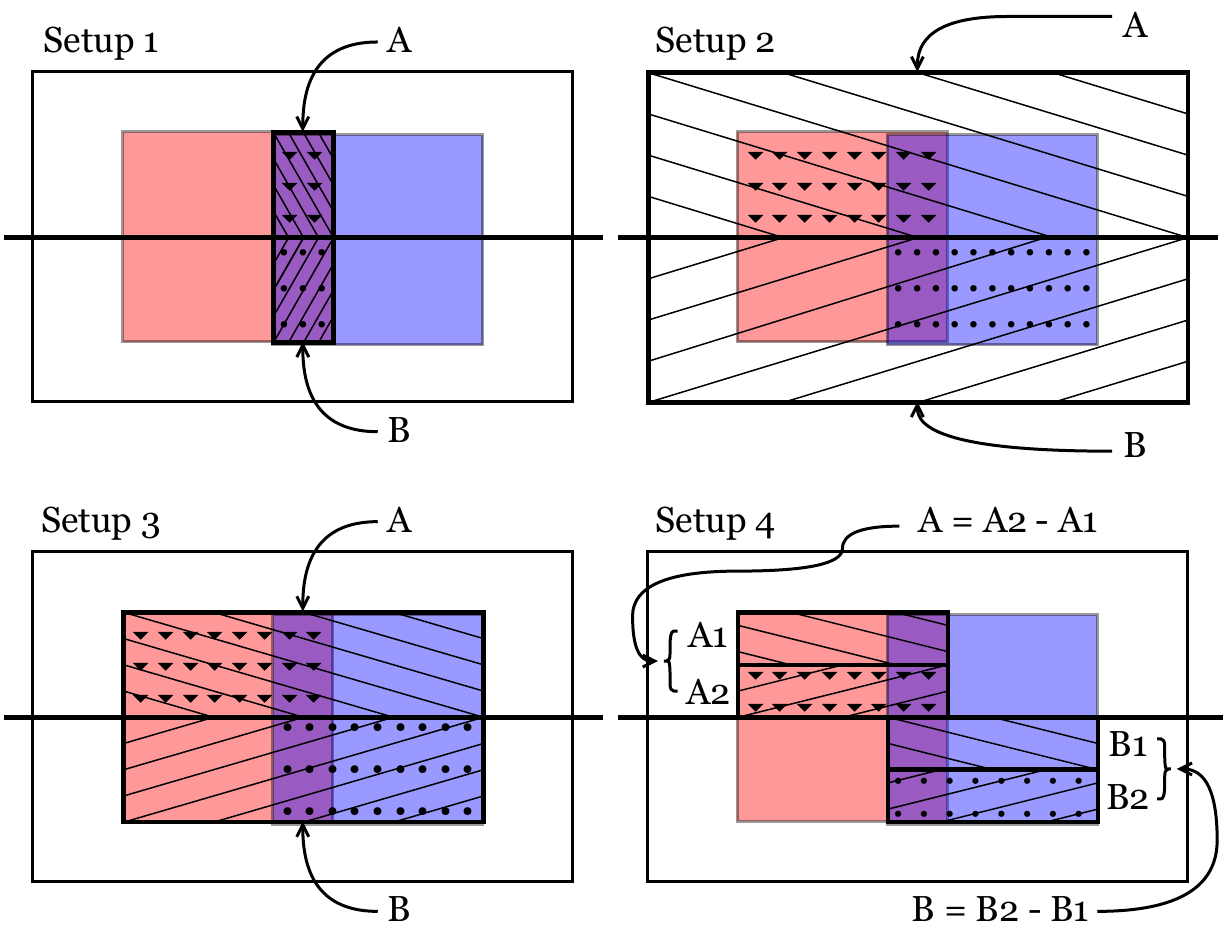}
\end{center}
\caption{Experiment setups overlaid on the user grouping Venn diagram in Figure~\ref{fig:ME_groups}. The hatched boxes indicate who are included in the analysis, and the downward triangles and dots indicate who are subjected to treatment prescribed under strategies 1 and 2 respectively. See Section~\ref{sec:experiment_setup} for a detailed description.}
\label{fig:ME_designs}
\end{figure}

Many experiment setups exist and are in use in different organizations. Here we introduce four common setups of various sophistication, which we also illustrate in Figure~\ref{fig:ME_designs}.

\paragraph{Setup 1 (Users in the intersection only)} The setup considers users who qualify for both strategies only. The said users are randomly split (usually 50/50) into two (analysis) groups $A$ and $B$, and are prescribed the treatment specified by strategies 1 and 2 respectively. The setup is easy to implement, though it is difficult to translate any learnings obtained from the setup to other user groups (e.g. those who qualify for one strategy only)~\cite{dmitriev17adirtydozen}.

\paragraph{Setup 2 (All samples)} The setup is a simple A/B test where it considers all users, regardless on whether they qualify for any strategy or not. The users are randomly split into two analysis groups $A$ and $B$, and are prescribed the treatment specified by strategy 1(2) if (i) they qualify under the strategy and (ii) they are in analysis group $A$($B$). This setup is easiest to implement but usually suffers severely from a dilution in metric~\cite{deng15diluted}.

\paragraph{Setup 3 (Qualified users only)} The setup is similar to Setup 2 except only those who qualified for at least one strategy (``triggered'' users in some literature~\cite{deng15diluted}) are included in the analysis groups. The setup sits between Setup 1 and Setup 2 in terms of user coverage, and has the advantage of capturing the most number of useful samples yet having the least metric dilution. However, the setup also prevents one from telling the incrementality of a strategy itself, but only the difference in incrementalities between two strategies.

\paragraph{Setup 4 (Dual control / multi-cell lift test)} As described in Section~\ref{sec:introduction}, the setup first split the users randomly into two randomization groups. For the first randomization group, we consider those who qualify for strategy 1, and split them into analysis groups $A1$ and $A2$. Group $A2$ receives the treatment prescribed under strategy~1, and group $A1$ acts as control. The incrementality for strategy~1 is then the difference in metric between groups $A2$ and $A1$. We apply the same process to the second randomization group, with strategy~2 and analysis groups $B1$ and $B2$ in place, and compare the incrementality for strategies 1 and 2. The setup allows one to obtain the incrementality of each individual strategy and minimizes metric dilution. Though it also leaves a number of samples unused and creates extra analysis groups, and hence generally suffers from a low test power~\cite{liu2018designing}.

\subsection{Evaluation criteria}
\label{sec:evaluation_criteria}
There are a number of considerations when one evaluates competing experiment setups. These include technical considerations such as the complexity of setting up the setups on an OCE framework, and business considerations such as whether the incrementality of individual strategies is required.

Here we focus on the statistical aspect and propose two evaluation criteria: (i) the actual average treatment effect size as presented by the two analysis groups in an experiment setup, and (ii) the sensitivity of the experiment represented by the minimum detectable effect (MDE) under a pre-specified test power. Both criteria are necessary as the former indicates whether a setup suffers from metric dilution, whereas the latter indicates whether the setup suffers from lack of power/sample size. An ideal setup should yield a high actual effect size and a high sensitivity (i.e. a low MDE),\footnote{We will use the terms ``high(er) sensitivity'' and ``low(er) MDE'' interchangeably.} though as we observe in the next section it is usually a trade-off.

We formally define the two evaluation criteria from first principles while introducing relevant notations along the way.
Let $A$ and $B$ be the two analysis groups in an experiment setup, with user responses randomly distributed with mean and variance $(\mu_A, \sigma^2_A)$ and $(\mu_B, \sigma^2_B)$ respectively.
We first recall that if there are sufficient samples, the sample mean of the two groups approximately follows the normal distribution by the Central Limit Theorem:
\begin{align}
    \bar{A} \overset{\textrm{approx.}}{\sim} \mathcal{N}\left(\mu_A, \nicefrac{\sigma^2_A}{n_A}\right), \quad
    \bar{B} \overset{\textrm{approx.}}{\sim} \mathcal{N}\left(\mu_B, \nicefrac{\sigma^2_B}{n_B}\right),
\end{align}
where $n_A$ and $n_B$ are the number of samples taken from $A$ and $B$ respectively. The difference in the sample means then also approximately follows a normal distribution:
\begin{align}
    \bar{D} \triangleq (\bar{B} - \bar{A}) \overset{\textrm{approx.}}{\sim} 
    \mathcal{N}\left(\Delta \triangleq \mu_B - \mu_A, 
    \,\sigma^2_{\bar{D}} \triangleq \nicefrac{\sigma^2_A}{n_A} + \nicefrac{\sigma^2_B}{n_B} \right).
    \label{eq:D_bar_distribution}
\end{align}
Here, $\Delta$ is the actual effect size that we are interested in.

The definition of the MDE $\theta^*$ requires a primer to the power of a statistical test. A common null hypothesis statistical test in personalization strategy experiments uses the two-tailed hypotheses $H_0: \Delta = 0$ and $H_1: \Delta \neq 0$, with the test statistic under $H_0$ being:
\begin{align}
    T \triangleq \nicefrac{\overline{D}}{\sigma_{\bar{D}}} \overset{\textrm{approx.}}{\sim} 
    \mathcal{N}\left(0, 1\right).
\end{align} 
We recall the null hypothesis will be rejected if $|T| > z_{1-\nicefrac{\alpha}{2}}$, where $\alpha$ is the significance level and $z_{1-\nicefrac{\alpha}{2}}$ is the~$1-\nicefrac{\alpha}{2}$ quantile of a standard normal. Under a \emph{specific} alternate hypothesis $\Delta = \theta$, the power is specified as
\begin{align}
    1 - \beta_\theta & \triangleq \textrm{Pr}\,\big(|T| > z_{1-\nicefrac{\alpha}{2}} \, |\, \Delta = \theta \big)
    \approx 1 - \Phi\big(z_{1 - \nicefrac{\alpha}{2}} - \nicefrac{|\theta|}{\sigma_{\bar{D}}}\big).
    \label{eq:beta_theta_approximation}
\end{align}
where $\Phi$ denotes the cumulative density function of a standard normal.\footnote{The approximation in Equation~\eqref{eq:beta_theta_approximation} is tight for experiment design purposes, where $\alpha < 0.2$ and $1 - \beta > 0.6$ for nearly all cases.}
To achieve a minimum test power $\pi_{\textrm{min}}$, we require that ${1 - \beta_\theta > \pi_{\textrm{min}}}$. Substituting Equation~\eqref{eq:beta_theta_approximation} into the inequality and rearranging to make $\theta$ the subject yields the effect sizes that the test will be able to detect with the specified power:
\begin{align}
    |\theta| > (z_{1 - \nicefrac{\alpha}{2}} - z_{1 - \pi_{\textrm{min}}})\,\sigma_{\bar{D}} \,.
    \label{eq:mde_ineq}
\end{align}
$\theta^*$ is then defined as the positive minimum $\theta$ that satisfies Inequality~\eqref{eq:mde_ineq}, i.e. that specified by the RHS of the inequality.

We finally define what it means to be better under these evaluation criteria. WLOG we assume the actual effect size of the two competing experiment setups are positive,\footnote{If both the actual effect sizes are negative, we simply swap the analysis groups. If the actual effect sizes are of opposite signs, it is likely an error.} and say
a setup $S$ is superior to another setup $R$ if, all else being equal,
\begin{enumerate}[leftmargin=*]
    \item[(i)] $S$ produces a higher actual effect size ($\Delta_S > \Delta_R$) \emph{and} a lower minimum detectable effect size ($\theta^*_S < \theta^*_R$), or
    \item[(ii)] The gain in actual effect is greater than the loss in sensitivity:
    \begin{align}
        \Delta_S - \Delta_R > \theta^*_S - \theta^*_R \,,
    \end{align}
    which means an actual effect still stands a higher chance to be observed under $S$.
\end{enumerate}


\section{Comparing experiment setups}
\label{sec:comparison}

Having described the evaluation framework above, in this section we use the framework to compare the common experiment setups described in Section~\ref{sec:experiment_setup}. We will first derive the actual effect size and MDE for each setup in Section~\ref{sec:actual_mde_size}, and using the result to create rule of thumbs on (i) whether diluting the metric by including users who qualify for neither strategies is beneficial (Section~\ref{sec:expt_comparison_dilution}) and (ii) if dual control is a better setup for personalization strategy experiments (Section~\ref{sec:expt_comparison_dual_control}), two questions that are often discussed among e-commerce and marketing-focused experimenters.
For brevity, we relegate most of the intermediate algebraic work when deriving the actual \& minimum detectable effect sizes, as well as the conditions that lead to a setup being superior, to our supplementary document.\footnotemark[2]

\subsection{Actual \& minimum detectable effect sizes}
\label{sec:actual_mde_size}

We first present the actual effect size and MDE of the four experiment setups. For each setup we first compute the sample size, mean response, and response variance in each analysis group, which arises as a mixture of user groups described in Section~\ref{sec:user_grouping}. For brevity, we only present the quantities for one analysis group per setup---expressions for other analysis group(s) can be easily obtained by substituting in the corresponding user groups. We then substitute the quantities computed into the definitions of~$\Delta$ (see Equation~\eqref{eq:D_bar_distribution}) and~$\theta^*$ (see Inequality~\eqref{eq:mde_ineq}) to obtain the setup-specific actual effect size and MDE. We assume all random splits are done 50/50 in these setups to maximize the test power.

\paragraph{Setup 1 (Users in the intersection only)}
The setup randomly splits user group 3 into two analysis groups, each with $\nicefrac{n_3}{2}$ samples. Users in analysis group $A$ are provided treatment under strategy~1, and hence the group's responses have a mean and variance of $(\mu_{I\phi}, \sigma^2_{I\phi})$.
The actual effect size and MDE for Setup 1 are hence:
\begin{align}
    \Delta_{S1} &= \mu_{I\psi} - \mu_{I\phi} , \label{eq:Delta_S1}\\
    \theta^*_{S1} &= (z_{1-\nicefrac{\alpha}{2}} - z_{1-\pi_{\textrm{min}}})
    \sqrt{2(\sigma^2_{I\phi} + \sigma^2_{I\psi})/n_3} .
    \label{eq:theta_star_S1}
\end{align}

\paragraph{Setup 2 (All samples)}
This setup also contains two analysis groups, $A$ and $B$, each taking half of the population (i.e. $(n_0 + n_1 + n_2 + n_3)/2$).
The mean response and response variance for groups $A$ and $B$ are the weighted mean response and response variance of the constituent user groups respectively. As we only provide treatment to those who qualify for strategy 1 in group $A$, and likewise for group $B$ with strategy 2, each user group will give different responses, e.g. for group $A$:
\begin{align}
    \mu_A = (n_0\mu_{C0} + n_1\mu_{I1} + n_2\mu_{C2} + n_3\mu_{I\phi}) / (n_0 + n_1 + n_2 + n_3), \label{eq:mu_A_S2}\\
    \sigma^2_A = (n_0\sigma^2_{C0} + n_1\sigma^2_{I1} + n_2\sigma^2_{C2} + n_3\sigma^2_{I\phi})/(n_0 + n_1 + n_2 + n_3). \label{eq:sigma_sq_A_S2}
\end{align}

Substituting the above (and that for group $B$) into the definitions of actual effect size and MDE we have:
\begin{align}
    \Delta_{S2} = \, &   \frac{n_1(\mu_{C1} - \mu_{I1}) + n_2(\mu_{I2} - \mu_{C2}) + n_3(\mu_{I\psi} - \mu_{I_\phi})}{n_0 + n_1 + n_2 + n_3},
    \label{eq:Delta_S2}\\
    \theta^*_{S2} = \, &  (z_{1-\nicefrac{\alpha}{2}} - z_{1-\pi_{\textrm{min}}})
    \sqrt{\frac{\begin{array}{l}
    2\big(n_0(2\sigma^2_{C0}) + n_1(\sigma^2_{I1} + \sigma^2_{C1}) + \\
    \quad n_2(\sigma^2_{C2} + \sigma^2_{I2}) + n_3(\sigma^2_{I\phi} + \sigma^2_{I\psi})\big)
    \end{array}
    }{(n_0 + n_1 + n_2 + n_3)^2}} .
    \label{eq:theta_star_S2}
\end{align}

\paragraph{Setup 3 (Qualified users only)}
The setup is very similar to Setup~2, with members from user group~0 excluded. This leads to both analysis groups having $(n_1 + n_2 + n_3)/2$ users.
The absence of group~0 users means they are not featured in the weighted mean response and response variance of the two analysis groups, e.g. for group A:
\begin{align}
    \mu_A = \frac{n_1\mu_{I1} + n_2\mu_{C2} + n_3\mu_{I\phi}}{n_1 + n_2 + n_3}, \;
    \sigma^2_A = \frac{n_1\sigma^2_{I1} + n_2\sigma^2_{C2} + n_3\sigma^2_{I\phi}}{n_1 + n_2 + n_3}.
    \label{eq:mu_A_sigma_sq_A_S3}
\end{align}

This leads to the following actual effect size and MDE for Setup~3:
\begin{align}
    \Delta_{S3} = &   \frac{n_1(\mu_{C1} - \mu_{I1}) + n_2(\mu_{I2} - \mu_{C2}) + n_3(\mu_{I\psi} - \mu_{I_\phi})}{n_1 + n_2 + n_3}, 
    \label{eq:Delta_S3}\\
    \theta^*_{S3} = &  (z_{1-\nicefrac{\alpha}{2}} \!-\! z_{1-\pi_{\textrm{min}}}) 
    \!\sqrt{\!\frac{2\!\left(n_1(\sigma^2_{I1} \!\!+\! \sigma^2_{C1}) \!+\! n_2(\sigma^2_{C2} \!\!+\! \sigma^2_{I2}) \!+\! n_3(\sigma^2_{I\phi} \!\!+\! \sigma^2_{I\psi})\!\right)\!}{(n_1 + n_2 + n_3)^2}}.
    \label{eq:theta_star_S3}
\end{align}


\paragraph{Setup 4 (Dual control)}
The setup is the odd one out as it has four analysis groups. Two of the analysis groups ($A1$ and $A2$) are drawn from those who qualified into strategy 1 and are allocated into the first randomization group, and the other two ($B1$ and $B2$) are drawn from those who are qualified into strategy 2 and are allocated into the second randomization group:
\begin{align}
    n_{A1} = n_{A2} = (n_1 + n_3)/4\,,\quad n_{B1} = n_{B2} = (n_2 + n_3)/4.
    \label{eq:n_S4}
\end{align}
The mean response and response variance for group $A1$ are:
\begin{align}
    \mu_{A1} = \frac{n_1\mu_{C1} + n_3\mu_{C3}}{n_1 + n_3},\;
    \sigma^2_{A1} = \frac{n_1\sigma^2_{C1} + n_3\sigma^2_{C3}}{n_1 + n_3}.
    \label{eq:mu_A1_sigma_sq_A1_S4}
\end{align}

As the setup takes the difference of differences in the metric (i.e. the difference between the mean response for groups $B2$ and $B1$, and the difference between the mean response for groups $A2$ and $A1$),\footnote{Not to be confused with the difference-in-differences method, which captures the metric for the control and treatment groups at both the beginning (pre-intervention) and the end (post-intervention) of the experiment. Here we only capture the metric once at the end of the experiment.} the actual effect size is as follows:
\begin{align}
    \Delta_{S4} & = (\mu_{B2} - \mu_{B1}) - (\mu_{A2} - \mu_{A1}) \nonumber\\
    & = \frac{n_2(\mu_{I2} \!-\! \mu_{C2}) + n_3(\mu_{I\psi} \!-\! \mu_{C3})}{n_2 + n_3} - \frac{n_2(\mu_{I1} \!-\! \mu_{C1}) + n_3(\mu_{I\phi} \!-\! \mu_{C3})}{n_1 + n_3} .
    \label{eq:Delta_S4}
\end{align}
The MDE for Setup 4 is similar to that specified in RHS of Inequality~\eqref{eq:mde_ineq}, albeit with more groups:
\begin{align}
    \theta^*_{S4} = \  & (z_{1-\nicefrac{\alpha}{2}} \!-\! z_{1-\pi_{\textrm{min}}})
    \sqrt{\nicefrac{\sigma^2_{A1}}{n_{A1}} + \nicefrac{\sigma^2_{A2}}{n_{A2}} + \nicefrac{\sigma^2_{B1}}{n_{B1}} + \nicefrac{\sigma^2_{B2}}{n_{B2}}} \nonumber\\
    = \,&  2 \cdot (z_{1-\nicefrac{\alpha}{2}} - z_{1-\pi_{\textrm{min}}}) \times \nonumber\\
    & \!\! \sqrt{\frac{n_1(\sigma^2_{C1} \!+\! \sigma^2_{I1}) \!+\! n_3(\sigma^2_{C3} \!+\! \sigma^2_{I\phi})}{(n_1 + n_3)^2} +
          \frac{n_2(\sigma^2_{C2} \!+\! \sigma^2_{I2}) \!+\! n_3(\sigma^2_{C3} \!+\! \sigma^2_{I\psi})}{(n_2 + n_3)^2}} .
    \label{eq:theta_star_S4}
\end{align}

\subsection{Is dilution always bad?}
\label{sec:expt_comparison_dilution}

The use of responses from users who do not qualify for any of the strategies we are comparing, an act known as metric dilution, has stirred countless debates in experimentation teams. On one hand, responses from these users make any treatment effect less pronounced by contributing exactly zero; on the other hand, it might be necessary as one does not know who actually qualify for a strategy~\cite{liu2018designing}, or it might be desirable as they can be leveraged to reduce the variance of the treatment effect estimator~\cite{deng15diluted}.

Here, we are interested in whether we should engage in the act of dilution given the assumed user responses prior to an experiment. This can be clarified by understanding the conditions where Setup~3 would emerge superior (as defined in Section~\ref{sec:evaluation_criteria}) to Setup~2. By inspecting Equations~\eqref{eq:Delta_S2} and~\eqref{eq:Delta_S3}, it is clear that $\Delta_{S3} > \Delta_{S2}$ if~${n_0 > 0}$. Thus, Setup~3 is superior to Setup~2 under the first criterion if $\theta^*_{S3} < \theta^*_{S2}$, which is the case if $\sigma^2_{C0}$, the metric variance of users who qualify for neither strategies, is large. This can be shown by substituting Equations~\eqref{eq:theta_star_S2} and~\eqref{eq:theta_star_S3} into the $\theta^*$-inequality and rearranging the terms to obtain:
\begin{align}
    \frac{\!\begin{array}{l}\left(n_1(\sigma^2_{I1} + \sigma^2_{C1}) + n_2(\sigma^2_{C2} + \sigma^2_{I2}) + n_3(\sigma^2_{I\phi} + \sigma^2_{I\psi})\right) \cdot \\  (n_0 + 2n_1 + 2n_2 + 2n_3)
    \end{array}\!}{2 (n_1 + n_2 + n_3)^2} < \sigma^2_{C0} \,.
    \label{eq:pse_setup23_criterion1}
\end{align}
If we assume the response variance are similar across groups with users who qualified for at least one strategy, i.e. $\sigma^2_{I1} \approx \sigma^2_{C1} \approx \cdots \approx \sigma^2_{I\psi} \approx \sigma^2_S$, Inequality~\eqref{eq:pse_setup23_criterion1} can then be simplified as
\begin{align}
    \sigma^2_S \left(\frac{n_0}{n_1 + n_2 + n_3} + 2\right) < \sigma^2_{C0} \,,
\end{align}
where it can be used to quickly determine if one should consider dilution at all.

If Inequality~\eqref{eq:pse_setup23_criterion1} is not true (i.e. $\theta^*_{S3} \geq \theta^*_{S2}$), we should then consider when the second criterion (i.e. $\Delta_{S3} - \Delta_{S2} > \theta^*_{S3} - \theta^*_{S2}$) is met. Writing
\begin{align}
    \eta & = n_1(\mu_{C1} - \mu_{I1}) + n_2(\mu_{I2} - \mu_{C2}) + n_3(\mu_{I\psi} - \mu_{I\phi}), \nonumber\\[-0.1em]
    \xi & = n_1(\sigma^2_{C1} + \sigma^2_{I1}) + n_2(\sigma^2_{I2} + \sigma^2_{C2}) + n_3(\sigma^2_{I\psi} + \sigma^2_{I\phi}), \textrm{and} \nonumber\\[-0.1em]
    z & = z_{1-\nicefrac{\alpha}{2}} - z_{1 - \pi_{\textrm{min}}}, \nonumber
\end{align}
we can substitute Equations~\eqref{eq:Delta_S2}, \eqref{eq:theta_star_S2}, \eqref{eq:Delta_S3}, and \eqref{eq:theta_star_S3} into the inequality and rearrange to obtain
\begin{align}
    \frac{n_1 + n_2 + n_3}{n_0}\sqrt{2n_0\sigma^2_{C0} + \xi} > 
    \frac{n_0 + n_1 + n_2 + n_3}{n_0}\sqrt{\xi} - \frac{\eta}{\sqrt{2}z}.
    \label{eq:pse_setup23_criterion2_initial}
\end{align}
As the LHS of Inequality~\eqref{eq:pse_setup23_criterion2_initial} is always positive, Setup 3 is superior if the RHS $\leq 0$. Noting 
\begin{align}
    \Delta_{S3} = \eta / (n_1 + n_2 + n_3) \,\textrm{ and }\,
    \theta^*_{S3} = \sqrt{2}\cdot z\cdot\sqrt{\xi} / (n_1 + n_2 + n_3), \nonumber
\end{align}
the trivial case is satisfied if
\begin{align}
\frac{n_0 + n_1 + n_2 + n_3}{n_0} \cdot \theta^*_{S3} \leq \Delta_{S3}.
    \label{eq:pse_setup23_criterion2_strong}
\end{align}

If the RHS of Inequality~\eqref{eq:pse_setup23_criterion2_initial} is positive, we can safely square both sides and use the identities for $\Delta_{S3}$ and $\theta^*_{S3}$ to get
\begin{align}
    \frac{2\sigma^2_{C0}}{n_0} >
    \frac{\left[\left(\theta^*_{S3} - \Delta_{S3} + \frac{n_1 + n_2 + n_3}{n_0} \theta^*_{S3}\right)^2 - \left(\frac{n_1 + n_2 + n_3}{n_0} \theta^*_{S3}\right)^2 \right]}{2z^2} .
    \label{eq:pse_setup23_criterion2_final}
\end{align}
As the LHS is always positive, the second criterion is met if
\begin{align}
    \theta^*_{S3} \leq \Delta_{S3}.
    \label{eq:pse_setup23_criterion2_weak}
\end{align}
Note this is a weaker, and thus more easily satisfiable condition than that introduced in Inequality~\eqref{eq:pse_setup23_criterion2_strong}.
This suggests an experiment setup is always superior to an diluted alternative  if the experiment is already adequately powered---introducing any dilution will simply make things worse.

Failing the condition in Inequality~\eqref{eq:pse_setup23_criterion2_weak}, we can always fall back to Inequality~\eqref{eq:pse_setup23_criterion2_final}. While the inequality operates in squared space, it is essentially comparing the standard error of user group~0 (LHS)---those who qualify for neither strategies---to the gap between the minimum detectable and actual effects ($\theta^*_{S3} - \Delta_{S3}$). The gap can be interpreted as the existing noise level, thus a higher standard error means mixing in group~0 users will introduce extra noise, and one is better off without them. Conversely, a smaller standard error means group~0 users can lower the noise level, i.e. stabilize the metric fluctuation, and one should take advantage of them.

To summarize, diluting a personalization strategies experiment setup is \emph{not} helpful if 
\begin{enumerate}[leftmargin=*]
    \item[(i)] Users who do not qualify for any strategies have a large metric variance (Inequality~\eqref{eq:pse_setup23_criterion1}), or
    \item[(ii)] The experiment is already adequately powered (Inequality~\eqref{eq:pse_setup23_criterion2_weak}).
\end{enumerate}
It could help if the experiment has not gained sufficient power yet and users who do not qualify for any strategy provide low-variance responses, such that they exhibit stabilizing effects when included into the analysis (complement of Inequality~\eqref{eq:pse_setup23_criterion2_final}).

\subsection{When is a dual-control more effective?}
\label{sec:expt_comparison_dual_control}
Often when advertisers compare two personalization strategies, the question on whether to use a dual control/multi-cell design comes up. Proponents of such approach celebrate its ability to tell a story by making the incrementality of an individual strategy available, while opponents voice concerns on the complexity in setting up the design. Here we are interested if Setup 4 (dual control) is superior to Setup 3 (a simple A/B test) from a power/detectable effect perspective, and if so, under what circumstances.

We first observe $\theta^*_{S4}$ > $\theta^*_{S3}$ is always true, and hence a dual control setup will never be superior to a simpler setup under the first criterion. This can be verified by substituting in Equations~\eqref{eq:theta_star_S4} and~\eqref{eq:theta_star_S3} and rearranging the terms to show the inequality is equivalent to
\begin{align}
    & \textstyle 2\Big(
      \frac{n_1}{(n_1 + n_3)^2} (\sigma^2_{C1} + \sigma^2_{I1}) +
      \frac{n_2}{(n_2 + n_3)^2} (\sigma^2_{C2} + \sigma^2_{I2}) +
      \nonumber\\ 
    & \textstyle \quad \frac{n_3}{(n_1 + n_3)^2} \sigma^2_{I\phi} +
      \frac{n_3}{(n_2 + n_3)^2} \sigma^2_{I\psi} +
      \left(\frac{n_3}{(n_1 + n_3)^2} + \frac{n_3}{(n_2 + n_3)^2}\right) \sigma^2_{C3} \Big) > \nonumber\\
    & \textstyle 
      \frac{n_1}{(n_1 + n_2 + n_3)^2} (\sigma^2_{C1} + \sigma^2_{I1}) +
      \frac{n_2}{(n_1 + n_2 + n_3)^2} (\sigma^2_{C2} + \sigma^2_{I2}) + \nonumber\\
    & \textstyle
      \quad \frac{n_3}{(n_1 + n_2 + n_3)^2} \sigma^2_{I\phi} + 
      \frac{n_3}{(n_1 + n_2 + n_3)^2} \sigma^2_{I\psi},
    \label{eq:pse_setup34_criterion1}
\end{align}
which is always true given the $n$s are non-negative and the $\sigma^2$s are positive: not only the coefficients of the $\sigma^2$-terms are larger on the LHS than their RHS counterparts, the LHS also carries an extra~$\sigma^2_{C3}$ term with non-negative coefficient and a factor of two.

Moving on to the second evaluation criterion, we recall that Setup 4 is superior if ${\Delta_{S4} - \Delta_{S3} > \theta^*_{S4} - \theta^*_{S3}}$, otherwise Setup 3 is superior under the same criterion. The full flexibility of the model can be seen by substituting Equations~\eqref{eq:Delta_S3}, \eqref{eq:theta_star_S3}, \eqref{eq:Delta_S4}, and~\eqref{eq:theta_star_S4} into the inequality and rearrange to obtain
\begin{align}
    &\frac{n_1 \frac{n_2(\mu_{I2}-\mu_{C2}) + n_3(\mu_{I\psi}-\mu_{C3})}{n_2 + n_3} -
           n_2 \frac{n_1(\mu_{I1}-\mu_{C1}) + n_3(\mu_{I\phi}-\mu_{C3})}{n_1 + n_3}}
           {\sqrt{n_1(\sigma^2_{C1} + \sigma^2_{I1}) +
                  n_2(\sigma^2_{C2} + \sigma^2_{I2}) +
                  n_3(\sigma^2_{I\phi} + \sigma^2_{I\psi})}} >
    \label{eq:pse_setup34_criterion2_full}\\
    & \sqrt{2}z \vast[
    \sqrt{
        2 \cdot 
        \frac{
            \begin{array}{l}
                (1 + \frac{n_2}{n_1 + n_3})^2 
                  [n_1(\sigma^2_{C1} + \sigma^2_{I1}) +
                  n_3(\sigma^2_{C3} + \sigma^2_{I\phi})] + \\
                \quad (1 + \frac{n_1}{n_2 + n_3})^2 
                  [n_2(\sigma^2_{C2} + \sigma^2_{I2}) +
                  n_3(\sigma^2_{C3} + \sigma^2_{I\psi})]
            \end{array}}
            {n_1(\sigma^2_{C1} + \sigma^2_{I1}) +
                n_2(\sigma^2_{C2} + \sigma^2_{I2}) +
                n_3(\sigma^2_{I\phi} + \sigma^2_{I\psi})}}
        - 1
    \vast],
    \nonumber
\end{align}
where $ z = z_{1-\nicefrac{\alpha}{2}} - z_{1 - \pi_{\textrm{min}}}$.

A key observation from inspecting Inequality~\eqref{eq:pse_setup34_criterion2_full} is that the LHS of the inequality scales along $O(\sqrt{n})$, where $n$ is the number of users, while the RHS remains a constant. This leads to the insight that Setup 4 is more likely to be superior if the $n$s are large. Here we assume the ratio $n_1:n_2:n_3$ remains unchanged when we scale the number of samples, an assumption that generally holds when an organization increases their reach while maintaining their user mix. It is worth pointing out that our claim is stronger than that in previous work --- we have shown that having a large user base not only fulfills the requirement of running a dual control experiment as described in~\cite{liu2018designing}, it also makes a dual control experiment a better setup than its simpler counterparts in terms of apparent and detectable effect sizes. 

The scaling relationship can be seen more clearly if we apply some simplifying assumptions to the $\sigma^2$- and $n$-terms. If we assume the response variances are similar across user groups (i.e. ${\sigma^2_{C1} \approx \sigma^2_{I1} \approx \cdots \approx \sigma^2_{I\psi} \approx \sigma^2_S}$), the RHS of Inequality~\eqref{eq:pse_setup34_criterion2_full} becomes
\begin{align}
    \sqrt{2} z \left[ \sqrt{\frac{n_1 + n_2 + n_3}{n_1 + n_3} + \frac{n_1 + n_2 + n_3}{n_2 + n_3}} - 1 \right],
    \label{eq:pse_setup34_criterion2_sigmaSimplified}
\end{align}
which remains a constant if the ratio $n_1:n_2:n_3$ remains unchanged. If we assume the number of users in groups 1, 2, 3 are similar (i.e. $n_1 = n_2 = n_3 = n$), the LHS of Inequality~\eqref{eq:pse_setup34_criterion2_full} becomes
\begin{align}
    \frac{\sqrt{n} \big((\mu_{I2} - \mu_{C2}) - (\mu_{I1} - \mu_{C1}) + \mu_{I\psi} - \mu_{I\phi}\big)}{2\sqrt{\sigma^2_{C1} + \sigma^2_{I1} + \sigma^2_{C2} + \sigma^2_{I2} + \sigma^2_{I\phi} + \sigma^2_{I\psi}}},
    \label{eq:pse_setup34_criterion2_nSimplified}
\end{align}
which clearly scales along $O(\sqrt{n})$.

We conclude the section by providing an indication on what a large $n$ may look like. If we assume both the response variances and the number of users are are similar across user groups, we can rearrange Inequality~\eqref{eq:pse_setup34_criterion2_full} to make $n$ the subject:
\begin{align}
    n > \left(2\sqrt{12}\left(\sqrt{6} - 1\right)z\right)^2 \frac{\sigma^2_S}{\Delta^2},
    \label{eq:pse_setup34_criterion2_bothSimplified}
\end{align}
where $\Delta = (\mu_{I2} - \mu_{C2}) - (\mu_{I1} - \mu_{C1}) + \mu_{I\psi} - \mu_{I\phi}$ is the difference in actual effect sizes between Setups 4 and 3. Under a 5\% significance level and 80\% power, the first coefficient amounts to around 791, which is roughly 50 times the coefficient one would use to determine the sample size of a simple A/B test~\cite{miller10hownot}. This suggests a dual control setup is perhaps a luxury accessible only to the largest advertising platforms and their top advertisers. For example, consider an experiment to optimize conversion rate where the baselines attain 20\% (hence having a variance of $0.2(1 - 0.2) = 0.16$). If there is a 2.5\% relative (i.e. 0.5\% absolute) effect between the competing strategies, the dual control setup will only be superior if there are $>5$M users in each user group. 

\section{Experiments}
\label{sec:experiments}
Having performed theoretical calculations for the actual and detectable effects and conditions where an experiment setup is superior to another, here we verify those calculations using simulation results.
We focus on the results presented in Section~\ref{sec:actual_mde_size}, as the rest of the results presented followed from those calculations.

In each experiment setup evaluation, we randomly select the value of the parameters (i.e. the $\mu$s, $\sigma^2$s, and $n$s), and take 1,000 actual effect samples, each by (i) sampling the responses from the user groups under the specified parameters, (ii) computing the mean for the analysis groups, and (iii) taking the difference of the means.

We also take 100 MDE samples in separate evaluations, each by (i) sampling a critical value under null hypothesis; (ii) computing the test power under a large number of possible effect sizes, each using the critical value and sampled metric means under the alternate hypothesis; and (iii) searching the effect size space for the value that gives the predefined power. As the power vs. effect size curve is noisy given the use of simulated power samples, we use the bisection algorithm provided by the \texttt{noisyopt} package to perform the search. The algorithm dynamically adjusts the number of samples taken from the same point on the curve to ensure the noise does not send us down the wrong search space. 

We expect the mean of the sampled actual effect and MDE to match the theoretical value.
To verify this, we perform~1,000 bootstrap resamplings on the samples obtained above to obtain an empirical bootstrap distribution of the sample mean in each evaluation. The $95\%$ bootstrap resampling confidence interval (BRCI) should then contain the theoretical mean $95\%$ of the times. 
The histogram of the percentile rank of the theoretical quantity in relation to the bootstrap samples across multiple evaluations should also follow a uniform distribution~\cite{talts2018validating}.

\begin{table}
\begin{center}
\begin{tabular}{|c|c|c|}
\hline
 & Actual effect size & Minimum detectable effect \\\hline
 Setup 1 & 1049/1099 (95.45\%) & 66/81 (81.48\%) \\\hline
 Setup 2 & 853/999 (85.38\%) & 87/106 (82.08\%) \\\hline
 Setup 3 & 922/1099 (83.89\%) & 93/116 (80.18\%) \\\hline
 Setup 4 & 240/333 (72.07\%) & 149/185 (80.54\%)\\
\hline
\end{tabular}
\end{center}
\vspace*{4pt}
\caption{Number of evaluations where the theoretical value of the quantities (columns) falls between the 95\% bootstrap confidence interval for each experiment setup (rows). See Section~\ref{sec:experiments} for a detailed description on the evaluations.}
\label{tab:experiment_in_BRCI_result}
\end{table}

The result is shown in Table~\ref{tab:experiment_in_BRCI_result}. One can observe that there are more evaluations having their theoretical quantity lying outside than the BRCI than expected. 
Upon further investigation, we observed a characteristic $\cup$-shape from the histograms of the percentile ranks for the actual effects. 
This suggests the bootstrap samples may be under-dispersed but otherwise centered on the theoretical quantities. 

We also observed the histograms for MDEs curving upward to the right, this suggests that the theoretical value is a slight overestimate (of $<1\%$ to the bootstrap mean in all cases). We believe this is likely due to a small bias in the bisection algorithm. The algorithm tests if the mean of the power samples is less than the target power to decide which half of the search space to continue along. Given we can bisect up to 10 times in each evaluation, it is likely to see a false positive even when we set the significance level for individual comparisons to 1\%. This leads to the algorithm favoring a smaller MDE sample. Having that said, since we have tested for a wide range of parameters and the overall bias is small, we are satisfied with the theoretical quantities for experiment design purposes.

\section{Conclusion}

We have addressed the problem of comparing experiment designs for personalization strategies by presenting an evaluation framework that allows experimenters to evaluate which experiment setup should be adopted given the situation. The flexible framework can be easily extended to compare setups that compare more than two strategies by adding more user groups (i.e. new sets to the Venn diagram in Figure~\ref{fig:ME_groups}). A new setup can also be incorporated quickly as it is essentially a different weighting of user group-scenario combinations shown in Table~\ref{tab:group_id}.
The framework also allows the development of simple rule of thumbs such as:
\begin{enumerate}[leftmargin=*]
    \item[(i)] Metric dilution should never be employed if the experiment already has sufficient power; though it can be useful if the experiment is under-powered and the non-qualifying users provide a ``stabilizing effect''; and
    \item[(ii)] A dual control setup is superior to simpler setups only if one has access to the user base of the largest organizations.
\end{enumerate}
We have validated the theoretical results via simulations, and made the code available\footnotemark[2] so that practitioners can benefit from the results immediately when designing their upcoming experiments.

\paragraph{Future Work}
So far we assume the responses from each user group-scenario combination are randomly distributed with the mean independent to the variance, with the evaluation criteria calculated assuming the metric (being the weighted sample mean of the responses) is approximately normally distributed under Central Limit Theorem. We are interested in whether the evaluation framework and its results still hold if we deviate from these assumptions, e.g. with binary responses (where the mean and variance correlate) and heavy-tailed distributed responses (where the sample mean converges to a normal distribution slowly).



\begin{acks}
The work is partially funded by the EPSRC CDT in Modern Statistics and Statistical Machine Learning at Imperial and Oxford (StatML.IO) and ASOS.com. The authors thank the anonymous reviewers for providing many improvements to the original manuscript.
\end{acks}

\bibliographystyle{ACM-Reference-Format}
\bibliography{experiment}

\clearpage

%
\appendix

\section*{Supplementary Document}

In this supplementary document, we revisit Section 3 of the paper ``An Evaluation Framework for Personalization Strategy Experiment Design'' by Liu and McCoy (2020), and expand on the intermediate algebraic work when deriving:
\begin{enumerate}[leftmargin=*]
    \item[(i)] The actual \& minimum detectable effect sizes, and
    \item[(ii)] The conditions that lead to a setup being superior.
\end{enumerate}
We will use the equation numbers featured in the original paper, and append letters for intermediate steps.

\section{Effect size of experiment setups}

We begin with the actual effect size and the MDE of the four experiment setups that are featured in Section~\ref{sec:actual_mde_size} of the paper. For each setup we first compute the sample size, mean response, and response variance in each analysis group (denoted $n_g$, $\mu_g$, and~$\sigma^2_g$ respectively for each analysis group $g$). These quantities arise as a mixture of user groups described in Section~\ref{sec:user_grouping} of the paper. We then substitute the quantities computed into the definitions of~$\Delta$ and~$\theta^*$:
\begin{align}
    \Delta & \triangleq \mu_B - \mu_A, \tag{see \eqref{eq:D_bar_distribution}}\\
    \theta^* & \triangleq (z_{1-\nicefrac{\alpha}{2}} - z_{1-\pi_{\textrm{min}}})\sigma_{\bar{D}}\,, \;\;
    \textrm{where } \sigma_{\bar{D}} = \nicefrac{\sigma^2_A}{n_A} + \nicefrac{\sigma^2_B}{n_B} \tag{see \eqref{eq:mde_ineq}}
\end{align}
and $z_q$ represents the $q^{\textrm{th}}$ quantile of a standard normal,
to obtain the setup-specific actual effect size and MDE for a setup with two analysis groups. For setups with more than two analysis groups, we will specify the actual and minimum detectable effect when we discuss specifics for each of the setups. We assume all random splits are done 50/50 in these setups to maximize the test power.

\paragraph{Setup 1 (Users in the intersection only)}
We recall the setup, which considers only users who qualify for both personalization strategies (i.e. the intersection), randomly splits user group 3 into two analysis groups, $A$ and $B$, each with the following number of samples:
\begin{align}
    n_A = n_B = \frac{n_3}{2}. \nonumber
\end{align}
Users in analysis group $A$ are provided treatment under strategy~1, and users in analysis group $B$ are provided treatment under strategy~2. This leads to the groups' responses having the following metric mean and variance: 
\begin{align}
    & \mu_A = \mu_{I\phi},\; \mu_B = \mu_{I\psi},\quad 
    \sigma^2_A = \sigma^2_{I\phi},\; \sigma^2_B = \sigma^2_{I\psi}. \nonumber
\end{align}

The actual effect size and MDE for Setup 1 are hence:
\begin{align}
    \Delta_{S1} &= \mu_{I\psi} - \mu_{I\phi}, \tag{\ref{eq:Delta_S1}} \\
    \theta^*_{S1} &= (z_{1-\nicefrac{\alpha}{2}} - z_{1-\pi_{\textrm{min}}})
    \sqrt{\frac{\sigma^2_{I\phi}}{\nicefrac{n_3}{2}} + \frac{\sigma^2_{I\psi}}{\nicefrac{n_3}{2}}}. \tag{\ref{eq:theta_star_S1}}
\end{align}

\paragraph{Setup 2 (All samples)}
The setup, which considers all users regardless of whether they qualify for any strategy or not, also contains two analysis groups, $A$ and $B$, each taking half of the population:
\begin{align}
    n_A = n_B = \frac{n_0 + n_1 + n_2 + n_3}{2} . \nonumber
\end{align}
The mean response and response variance for groups $A$ and $B$ are the weighted mean response and response variance of the constituent user groups respectively, weighted by the constituent groups' size. As we only provide treatment to those who qualify for strategy~1 in group~A, and those who qualify for strategy~2 in group~B, this leads to different responses in different constituent user groups:
\begin{align}
    \mu_A = \frac{n_0\mu_{C0} + n_1\mu_{I1} + n_2\mu_{C2} + n_3\mu_{I\phi}}{n_0 + n_1 + n_2 + n_3}, \tag{\ref{eq:mu_A_S2}}\\
    \mu_B = \frac{n_0\mu_{C0} + n_1\mu_{C1} + n_2\mu_{I2} + n_3\mu_{I\psi}}{n_0 + n_1 + n_2 + n_3}; \nonumber\\
    \sigma^2_A = \frac{n_0\sigma^2_{C0} + n_1\sigma^2_{I1} + n_2\sigma^2_{C2} + n_3\sigma^2_{I\phi}}{n_0 + n_1 + n_2 + n_3}, \tag{\ref{eq:sigma_sq_A_S2}}\\
    \sigma^2_B = \frac{n_0\sigma^2_{C0} + n_1\sigma^2_{C1} + n_2\sigma^2_{I2} + n_3\sigma^2_{I\psi}}{n_0 + n_1 + n_2 + n_3}. \nonumber
\end{align}

Substituting the above into the definitions of actual effect size and MDE and simplifying the resultant expressions we have:
\begin{align}
    \Delta_{S2} = \, &   \frac{n_1(\mu_{C1} - \mu_{I1}) + n_2(\mu_{I2} - \mu_{C2}) + n_3(\mu_{I\psi} - \mu_{I_\phi})}{n_0 + n_1 + n_2 + n_3}, \tag{\ref{eq:Delta_S2}} \\
    \theta^*_{S2} = \, &  (z_{1-\nicefrac{\alpha}{2}} - z_{1-\pi_{\textrm{min}}})
    \sqrt{\frac{\begin{array}{l}
    2\big(n_0(2\sigma^2_{C0}) + n_1(\sigma^2_{I1} + \sigma^2_{C1}) + \\
    \quad n_2(\sigma^2_{C2} + \sigma^2_{I2}) + n_3(\sigma^2_{I\phi} + \sigma^2_{I\psi})\big)
    \end{array}
    }{(n_0 + n_1 + n_2 + n_3)^2}} . \tag{\ref{eq:theta_star_S2}}
\end{align}

\paragraph{Setup 3 (Qualified users only)}
The setup is very similar to Setup~2, with members from user group~0 excluded:
\begin{align}
    n_A = n_B = \frac{n_1 + n_2 + n_3}{2}. \nonumber
\end{align}
The absence of members from user group 0 means they are not featured in the weighted mean response and response variance of the two analysis groups:
\begin{align}
    \mu_A = \frac{n_1\mu_{I1} + n_2\mu_{C2} + n_3\mu_{I\phi}}{n_1 + n_2 + n_3}, \;
    \mu_B = \frac{n_1\mu_{C1} + n_2\mu_{I2} + n_3\mu_{I\psi}}{n_1 + n_2 + n_3}; \nonumber\\
    \sigma^2_A = \frac{n_1\sigma^2_{I1} + n_2\sigma^2_{C2} + n_3\sigma^2_{I\phi}}{n_1 + n_2 + n_3}, \;
    \sigma^2_B = \frac{n_1\sigma^2_{C1} + n_2\sigma^2_{I2} + n_3\sigma^2_{I\psi}}{n_1 + n_2 + n_3}. \tag{\ref{eq:mu_A_sigma_sq_A_S3}}
\end{align}

This lead to the following actual effect size and MDE for Setup 3:
\begin{align}
    \Delta_{S3} = &   \frac{n_1(\mu_{C1} - \mu_{I1}) + n_2(\mu_{I2} - \mu_{C2}) + n_3(\mu_{I\psi} - \mu_{I_\phi})}{n_1 + n_2 + n_3}, 
    \tag{\ref{eq:Delta_S3}}\\
    \theta^*_{S3} = &  (z_{1-\nicefrac{\alpha}{2}} \!-\! z_{1-\pi_{\textrm{min}}}) 
    \!\sqrt{\!\frac{2\!\left(n_1(\sigma^2_{I1} \!\!+\! \sigma^2_{C1}) \!+\! n_2(\sigma^2_{C2} \!\!+\! \sigma^2_{I2}) \!+\! n_3(\sigma^2_{I\phi} \!\!+\! \sigma^2_{I\psi})\!\right)\!}{(n_1 + n_2 + n_3)^2}}.
    \tag{\ref{eq:theta_star_S3}}
\end{align}

\paragraph{Setup 4 (Dual control)}
Setup 4 is unique amongst the experiment setups introduced as it has four analysis groups. Two of the analysis groups are drawn from those who qualified into strategy 1 and are allocated into the first half, and the other two are drawn from those who are qualified into strategy 2 and are allocated into the second half:
\begin{align}
    n_{A1} = n_{A2} = \frac{n_1 + n_3}{4},\; n_{B1} = n_{B2} = \frac{n_2 + n_3}{4}.
    \tag{\ref{eq:n_S4}}
\end{align}
The mean response and response variance of each analysis group are the weighted metric mean response and response variance of the user groups involved respectively:
\begin{align}
    \mu_{A1} = \frac{n_1\mu_{C1} + n_3\mu_{C3}}{n_1 + n_3},\, 
    \mu_{A2} = \frac{n_1\mu_{I1} + n_3\mu_{I\phi}}{n_1 + n_3}, \nonumber\\
    \mu_{B1} = \frac{n_2\mu_{C2} + n_3\mu_{C3}}{n_2 + n_3},\,
    \mu_{B2} = \frac{n_2\mu_{I2} + n_3\mu_{I\psi}}{n_2 + n_3}; \nonumber\\
    \sigma^2_{A1} = \frac{n_1\sigma^2_{C1} + n_3\sigma^2_{C3}}{n_1 + n_3},\,
    \sigma^2_{A2} = \frac{n_1\sigma^2_{I1} + n_3\sigma^2_{I\phi}}{n_1 + n_3}, \nonumber\\
    \sigma^2_{B1} = \frac{n_2\sigma^2_{C2} + n_3\sigma^2_{C3}}{n_2 + n_3},\,
    \sigma^2_{B2} = \frac{n_2\sigma^2_{I2} + n_3\sigma^2_{I\psi}}{n_2 + n_3}; \tag{\ref{eq:mu_A1_sigma_sq_A1_S4}}
\end{align}

As the setup takes the difference of differences (i.e. the difference of groups $B2$ and $B1$, and the difference of groups $A2$ and $A1$), the actual effect size are specified, post-simplification, as follows:
\begin{align}
    \Delta_{S4} & = (\mu_{B2} - \mu_{B1}) - (\mu_{A2} - \mu_{A1}) \nonumber\\
    & = \frac{n_2(\mu_{I2} \!-\! \mu_{C2}) + n_3(\mu_{I\psi} \!-\! \mu_{C3})}{n_2 + n_3} - \frac{n_1(\mu_{I1} \!-\! \mu_{C1}) + n_3(\mu_{I\phi} \!-\! \mu_{C3})}{n_1 + n_3} .
    \tag{\ref{eq:Delta_S4}}
\end{align}
The MDE for Setup 4 is similar to that specified in the RHS of Inequality~\eqref{eq:mde_ineq}, albeit with more groups:
\begin{align}
    \theta^*_{S4} = \  & (z_{1-\nicefrac{\alpha}{2}} \!-\! z_{1-\pi_{\textrm{min}}})
    \sqrt{\nicefrac{\sigma^2_{A1}}{n_{A1}} + \nicefrac{\sigma^2_{A2}}{n_{A2}} + \nicefrac{\sigma^2_{B1}}{n_{B1}} + \nicefrac{\sigma^2_{B2}}{n_{B2}}} \nonumber\\
    = \,&  2 \cdot (z_{1-\nicefrac{\alpha}{2}} - z_{1-\pi_{\textrm{min}}}) \times \nonumber\\
    & \!\! \sqrt{\frac{n_1(\sigma^2_{C1} \!+\! \sigma^2_{I1}) \!+\! n_3(\sigma^2_{C3} \!+\! \sigma^2_{I\phi})}{(n_1 + n_3)^2} +
          \frac{n_2(\sigma^2_{C2} \!+\! \sigma^2_{I2}) \!+\! n_3(\sigma^2_{C3} \!+\! \sigma^2_{I\psi})}{(n_2 + n_3)^2}} .
    \tag{\ref{eq:theta_star_S4}}
\end{align}

\section{Metric dilution}
We then expand the calculations in Section~\ref{sec:expt_comparison_dilution} of the paper, where we discuss the conditions where an experiment setup with metric dilution (Setup~2) will emerge superior to one without metric dilution (Setup~3), and vice versa.

\subsection{The first criterion}

We first show $\theta^*_{S3} < \theta^*_{S2}$, the condition which will lead to Setup~3 being superior to Setup~2 under the first criterion, is equivalent to 
\begin{align}
    \frac{\!\begin{array}{l}\big(n_1(\sigma^2_{I1} + \sigma^2_{C1}) + n_2(\sigma^2_{C2} + \sigma^2_{I2}) + n_3(\sigma^2_{I\phi} + \sigma^2_{I\psi})\big) \cdot \\  (n_0 + 2n_1 + 2n_2 + 2n_3)
    \end{array}\!}{2 (n_1 + n_2 + n_3)^2} < \sigma^2_{C0}.
    \tag{\ref{eq:pse_setup23_criterion1}}
\end{align}

We start by substituting the expressions for $\theta^*_{S2}$ (Equation~\eqref{eq:theta_star_S2}) and $\theta^*_{S3}$ (Equation~\eqref{eq:theta_star_S3}) into the inequality $\theta^*_{S3} < \theta^*_{S2}$ to obtain
\begin{align}
    & (z_{1-\nicefrac{\alpha}{2}} \!-\! z_{1-\pi_{\textrm{min}}}) 
    \!\sqrt{\!\frac{2\big(n_1(\sigma^2_{I1} \!\!+\! \sigma^2_{C1}) \!+\! n_2(\sigma^2_{C2} \!\!+\! \sigma^2_{I2}) \!+\! n_3(\sigma^2_{I\phi} \!\!+\! \sigma^2_{I\psi})\!\big)\!}{(n_1 + n_2 + n_3)^2}} \nonumber\\
    <\, & (z_{1-\nicefrac{\alpha}{2}} - z_{1-\pi_{\textrm{min}}})
    \sqrt{\frac{\begin{array}{l}
    2\big(n_0(2\sigma^2_{C0}) + n_1(\sigma^2_{I1} + \sigma^2_{C1}) + \\
    \quad n_2(\sigma^2_{C2} + \sigma^2_{I2}) + n_3(\sigma^2_{I\phi} + \sigma^2_{I\psi})\big)
    \end{array}
    }{(n_0 + n_1 + n_2 + n_3)^2}}.
    \tag{\ref{eq:pse_setup23_criterion1}a}
\end{align}
Canceling the $z_{1-\nicefrac{\alpha}{2}} - z_{1 - \pi_{\min}}$ and $\sqrt{2}$ terms on both sides, and then squaring them yields
\begin{align}
    & \frac{n_1(\sigma^2_{I1} + \sigma^2_{C1}) + n_2(\sigma^2_{C2} + \sigma^2_{I2}) + n_3(\sigma^2_{I\phi} + \sigma^2_{I\psi})}{(n_1 + n_2 + n_3)^2} \nonumber\\
    <\, & \frac{
    n_0(2\sigma^2_{C0}) + n_1(\sigma^2_{I1} + \sigma^2_{C1}) +
    n_2(\sigma^2_{C2} + \sigma^2_{I2}) + n_3(\sigma^2_{I\phi} + \sigma^2_{I\psi})
    }{(n_0 + n_1 + n_2 + n_3)^2}.
    \tag{\ref{eq:pse_setup23_criterion1}b}
\end{align}

We then write $\xi = n_1(\sigma^2_{I1} + \sigma^2_{C1}) + n_2(\sigma^2_{C2} + \sigma^2_{I2}) + n_3(\sigma^2_{I\phi} + \sigma^2_{I\psi})$ and move the $\xi$ terms on the RHS to the LHS:
\begin{align}
    \xi\left(\frac{1}{(n_1 + n_2 + n_3)^2} - \frac{1}{(n_0 + n_1 + n_2 + n_3)^2}\right) < \frac{n_0(2\sigma^2_{C0})}{(n_0 + n_1 + n_2 + n_3)^2}.
    \tag{\ref{eq:pse_setup23_criterion1}c}
    \label{eq:pse_setup23_criterion1c}
\end{align}
As the partial fractions can be consolidated as
\begin{align}
    & \frac{1}{(n_1 + n_2 + n_3)^2} - \frac{1}{(n_0 + n_1 + n_2 + n_3)^2} \nonumber\\
    =\, & \frac{(n_0 + n_1 + n_2 + n_3)^2 - (n_1 + n_2 + n_3)^2}{(n_1 + n_2 + n_3)^2(n_0 + n_1 + n_2 + n_3)^2} \nonumber\\
    =\, & \frac{(n_0 + 2n_1 + 2n_2 + 2n_3)n_0}{(n_1 + n_2 + n_3)^2(n_0 + n_1 + n_2 + n_3)^2}\,,
    \tag{\ref{eq:pse_setup23_criterion1}d}
\end{align}
where the second step utilizes the identity $a^2 - b^2 = (a+b)(a-b)$, Inequality~\eqref{eq:pse_setup23_criterion1c} can be written as
\begin{align}
    \xi\left(\frac{(n_0 + 2n_1 + 2n_2 + 2n_3)n_0}{(n_1 + n_2 + n_3)^2(n_0 + n_1 + n_2 + n_3)^2}\right) < \frac{n_0(2\sigma^2_{C0})}{(n_0 + n_1 + n_2 + n_3)^2}.
    \tag{\ref{eq:pse_setup23_criterion1}e}
    \label{eq:pse_setup23_criterion1e}
\end{align}
We finally cancel the $n_0$ and $(n_0+n_1+n_2+n_3)^2$ terms on both sides of Inequality~\eqref{eq:pse_setup23_criterion1e}, move the factor of two to the LHS, and write $\xi$ in its full form to arrive at Inequality~\eqref{eq:pse_setup23_criterion1}:
\begin{align}
    \frac{\!\begin{array}{l}\big(n_1(\sigma^2_{I1} + \sigma^2_{C1}) + n_2(\sigma^2_{C2} + \sigma^2_{I2}) + n_3(\sigma^2_{I\phi} + \sigma^2_{I\psi})\big) \cdot \\  (n_0 + 2n_1 + 2n_2 + 2n_3)
    \end{array}\!}{2 (n_1 + n_2 + n_3)^2} < \sigma^2_{C0}.
    \nonumber
\end{align}

\subsection{The second criterion}
In the case where Inequality~\eqref{eq:pse_setup23_criterion1} does not hold, we consider when Setup~3 will emerge superior to Setup~2 under the second criterion:
\begin{align}
    \Delta_{S3} - \Delta_{S2} > \theta^*_{S3} - \theta^*_{S2}. \nonumber
\end{align}
If this inequality does not hold either (and both sides are not equal), we consider Setup~2 as superior to Setup~3 under the same criterion as the following holds:
\begin{align}
    \Delta_{S3} - \Delta_{S2} < \theta^*_{S3} - \theta^*_{S2} 
    \;\;\iff\;\;  \Delta_{S2} - \Delta_{S3} > \theta^*_{S2} - \theta^*_{S3}.
    \nonumber
\end{align}

\paragraph{The master inequality}
We first show that the inequality $\Delta_{S3} - \Delta_{S2} > \theta^*_{S3} - \theta^*_{S2}$ is equivalent to
\begin{align}
    \frac{n_1 + n_2 + n_3}{n_0}\sqrt{2n_0\sigma^2_{C0} + \xi} > 
    \frac{n_0 + n_1 + n_2 + n_3}{n_0}\sqrt{\xi} - \frac{\eta}{\sqrt{2}z}
    \tag{\ref{eq:pse_setup23_criterion2_initial}},
\end{align}
where
\begin{align}
    \eta & = n_1(\mu_{C1} - \mu_{I1}) + n_2(\mu_{I2} - \mu_{C2}) + n_3(\mu_{I\psi} - \mu_{I\phi}), \nonumber\\[-0.1em]
    \xi & = n_1(\sigma^2_{C1} + \sigma^2_{I1}) + n_2(\sigma^2_{I2} + \sigma^2_{C2}) + n_3(\sigma^2_{I\psi} + \sigma^2_{I\phi}), \textrm{and} \nonumber\\[-0.1em]
    z & = z_{1-\nicefrac{\alpha}{2}} - z_{1 - \pi_{\textrm{min}}}\,. \nonumber
\end{align}
Writing $\eta$, $\xi$, and $z$ as shown above, we substitute in the expressions for $\Delta_{S2}$, $\theta^*_{S2}$, $\Delta_{S3}$, and $\theta^*_{S3}$ (Equations~\eqref{eq:Delta_S2}, \eqref{eq:theta_star_S2}, \eqref{eq:Delta_S3}, and \eqref{eq:theta_star_S3} respectively) into the initial inequality to obtain
\begin{align}
    & \frac{\eta}{n_1 + n_2 + n_3} - 
    \frac{\eta}{n_0 + n_1 + n_2 + n_3} \nonumber\\
    >\, & z\sqrt{\frac{2\xi}{(n_1 + n_2 + n_3)^2}} - 
    z\sqrt{\frac{2\big(n_0(2\sigma^2_{C0})+ \xi\big)}{(n_0 + n_1 + n_2 + n_3)^2}} .
    \tag{\ref{eq:pse_setup23_criterion2_initial}a}
\end{align}
Pulling out the common factors on each side we have
\begin{align}
    & \eta\left(\frac{1}{n_1 + n_2 + n_3} - 
    \frac{1}{n_0 + n_1 + n_2 + n_3}\right) \nonumber\\
    >\, & \sqrt{2} z \Bigg(\frac{\sqrt{\xi}}{n_1 + n_2 + n_3} - \frac{\sqrt{2n_0\sigma^2_{C0}+ \xi}}{n_0 + n_1 + n_2 + n_3}\Bigg).
    \tag{\ref{eq:pse_setup23_criterion2_initial}b}
\end{align}
Writing the partial fraction on the LHS of
Inequality~(\ref{eq:pse_setup23_criterion2_initial}b) as a composite fraction we have
\begin{align}
    & \eta\left(\frac{n_0}{(n_1+n_2+n_3)(n_0+n_1+n_2+n_3)}\right) \nonumber\\
    >\, & \sqrt{2} z \Bigg(\frac{\sqrt{\xi}}{n_1 + n_2 + n_3} - \frac{\sqrt{2n_0\sigma^2_{C0}+ \xi}}{n_0 + n_1 + n_2 + n_3}\Bigg).
    \tag{\ref{eq:pse_setup23_criterion2_initial}c}
\end{align}
We then move the composite fraction to the RHS and the $\sqrt{2}z$ term to the LHS:
\begin{align}
    \frac{\eta}{\sqrt{2}z} >
    \frac{\begin{array}{ll}(n_1+n_2+n_3)\cdot\\(n_0+n_1+n_2+n_3)\end{array}}{n_0}
    \Bigg(\frac{\sqrt{\xi}}{n_1 + n_2 + n_3} - \frac{\sqrt{2n_0\sigma^2_{C0}+ \xi}}{n_0 + n_1 + n_2 + n_3}\Bigg),
    \tag{\ref{eq:pse_setup23_criterion2_initial}d}
\end{align}
and expand the brackets, canceling terms that appear on both sides of the fractions in the RHS:
\begin{align}
    \frac{\eta}{\sqrt{2}z} >
    \frac{n_0 + n_1 + n_2 + n_3}{n_0} \sqrt{\xi} - \frac{n_1 + n_2 + n_3}{n_0} \sqrt{2n_0\sigma^2_{C0}+ \xi} \;.
    \tag{\ref{eq:pse_setup23_criterion2_initial}e}
\end{align}
Finally, we swap the position of the leftmost term with that of the rightmost term in Inequality~(\ref{eq:pse_setup23_criterion2_initial}e) to arrive at Inequality~\eqref{eq:pse_setup23_criterion2_initial}:
\begin{align}
    \frac{n_1 + n_2 + n_3}{n_0}\sqrt{2n_0\sigma^2_{C0} + \xi} > 
    \frac{n_0 + n_1 + n_2 + n_3}{n_0}\sqrt{\xi} - \frac{\eta}{\sqrt{2}z}.
    \nonumber
\end{align}

\paragraph{The trivial case: RHS $\leq 0$}
We first observe that the LHS of Inequality~\eqref{eq:pse_setup23_criterion2_initial} is always positive, and hence
the inequality trivially holds if the RHS is non-positive.
Here we show RHS $\leq 0$ is equivalent to
\begin{align}
    \frac{n_0 + n_1 + n_2 + n_3}{n_0}\theta^*_{S3} \leq \Delta_{S3}.
    \tag{\ref{eq:pse_setup23_criterion2_strong}}
\end{align}
The can be done by writing RHS $\leq 0$ in full:
\begin{align}
    \frac{n_0 + n_1 + n_2 + n_3}{n_0}\sqrt{\xi} - \frac{\eta}{\sqrt{2}z} \leq 0, \tag{\ref{eq:pse_setup23_criterion2_strong}a}
\end{align}
and moving the second term on the LHS to the RHS:
\begin{align}
    \frac{n_0 + n_1 + n_2 + n_3}{n_0}\sqrt{\xi} \leq \frac{\eta}{\sqrt{2}z}. \tag{\ref{eq:pse_setup23_criterion2_strong}b}
\end{align}
We then add a factor of $\sqrt{2}z / (n_1 + n_2 + n_3)$ on both sides to get
\begin{align}
    \frac{n_0 + n_1 + n_2 + n_3}{n_0}\frac{\sqrt{\xi}\sqrt{2}z}{n_1+n_2+n_3} \leq \frac{\eta}{n_1+n_2+n_3}. \tag{\ref{eq:pse_setup23_criterion2_strong}c}
    \label{eq:23c}
\end{align}
Noting from Equations \eqref{eq:Delta_S3} and \eqref{eq:theta_star_S3} that
\begin{align}
    \Delta_{S3} = \frac{\eta}{n_1 + n_2 + n_3} \,\textrm{ and }\,
    \theta^*_{S3} = \frac{\sqrt{2}\cdot z\cdot\sqrt{\xi}}{n_1 + n_2 + n_3}, \nonumber
\end{align}
we finally replace the terms in Inequality~\eqref{eq:23c} with $\Delta_{S3}$ and $\theta^*_{S3}$ to arrive at Inequality~\eqref{eq:pse_setup23_criterion2_strong}:
\begin{align}
    \frac{n_0 + n_1 + n_2 + n_3}{n_0}\theta^*_{S3} \leq \Delta_{S3}.
    \nonumber
\end{align}

\paragraph{The non-trivial case: RHS $>0$}
We then tackle the case where the RHS of the master inequality (Inequality~\eqref{eq:pse_setup23_criterion2_initial}) is greater than zero. We show in this non-trivial case, Inequality~\eqref{eq:pse_setup23_criterion2_initial} is equivalent to
\begin{align}
    \frac{2\sigma^2_{C0}}{n_0} >
    \frac{\left(\theta^*_{S3} - \Delta_{S3} + \frac{n_1 + n_2 + n_3}{n_0} \theta^*_{S3}\right)^2 - \left(\frac{n_1 + n_2 + n_3}{n_0} \theta^*_{S3}\right)^2}{2z^2} .
    \tag{\ref{eq:pse_setup23_criterion2_final}}
\end{align}
We first multiply both sides of Inequality~\eqref{eq:pse_setup23_criterion2_initial} with the fraction $n_0\sqrt{2}z / (n_1 + n_2 + n_3)$ to get
\begin{align}
    & \frac{n_1 + n_2 + n_3}{n_0}\sqrt{2n_0\sigma^2_{C0} + \xi}
    \frac{n_0\sqrt{2}z}{n_1 + n_2 + n_3} \nonumber\\
    > &  
    \left(\frac{n_0 + n_1 + n_2 + n_3}{n_0}\sqrt{\xi} - \frac{\eta}{\sqrt{2}z}\right)
    \frac{n_0\sqrt{2}z}{n_1 + n_2 + n_3} \;.
    \tag{\ref{eq:pse_setup23_criterion2_final}a}
\end{align}
Canceling terms on both sides of the fractions we have
\begin{align}
    & \sqrt{2n_0\sigma^2_{C0} + \xi} \sqrt{2}z \nonumber \\
    >\, & (n_0 + n_1 + n_2 + n_3) \frac{\sqrt{\xi}\sqrt{2}z}{n_1 + n_2 + n_3} - n_0 \frac{\eta}{n_1 + n_2 + n_3} \;. \tag{\ref{eq:pse_setup23_criterion2_final}b}
\end{align}
Again noting the identities for $\Delta_{S3}$ and $\theta^*_{S3}$ stated above, we can replace the fractions on the RHS and obtain
\begin{align}
    \sqrt{2n_0\sigma^2_{C0} + \xi} \sqrt{2}z 
    > (n_0 + n_1 + n_2 + n_3) \theta^*_{S3} - n_0 \Delta_{S3}.
    \tag{\ref{eq:pse_setup23_criterion2_final}c}
    \label{eq:pse_setup23_criterion2_final_c}
\end{align}

We then square both sides of Inequality~\eqref{eq:pse_setup23_criterion2_final_c}
and move the $2z^2$ term to the RHS:
\begin{align}
    2n_0\sigma^2_{C0} + \xi
    > \frac{\big((n_0 + n_1 + n_2 + n_3) \theta^*_{S3} - n_0 \Delta_{S3}\big)^2}{2z^2} .
    \tag{\ref{eq:pse_setup23_criterion2_final}d}
    \label{eq:pse_setup23_criterion2_final_d}
\end{align}
Note the squaring still allows the implication to go both ways as both sides of Inequality~\eqref{eq:pse_setup23_criterion2_final_c} are positive.
Based on the identity for $\theta^*_{S3}$, we observe $\xi$ can also be written as
\begin{align}
    \xi = \frac{(n_1 + n_2 + n_3)^2 (\theta^*_{S3})^2}{2z^2} .
    \tag{\ref{eq:pse_setup23_criterion2_final}e}
    \label{eq:pse_setup23_criterion2_final_e}
\end{align}
Thus, we can group all terms with a $2z^2$ denominator by moving $\xi$ in Inequality~\eqref{eq:pse_setup23_criterion2_final_d} to the RHS and substituting Equation~\eqref{eq:pse_setup23_criterion2_final_e} into the resultant inequality:
\begin{align}
    2n_0\sigma^2_{C0}
    > \frac{\big((n_0 + n_1 + n_2 + n_3) \theta^*_{S3} - n_0 \Delta_{S3}\big)^2 \!-\! \big((n_1 + n_2 + n_3) \theta^*_{S3}\big)^2}{2z^2} .
    \tag{\ref{eq:pse_setup23_criterion2_final}f}
    \label{eq:pse_setup23_criterion2_final_f}
\end{align}

We finally normalize the inequality to one with unit~$\Delta_{S3}$ and~$\theta^*_{S3}$ to enable effective comparison. 
We divide both sides of Inequality~\eqref{eq:pse_setup23_criterion2_final_f} by ${n_0}^2$:
\begin{align}
    \frac{2\sigma^2_{C0}}{n_0}
    > \frac{\Big(\frac{n_0 + n_1 + n_2 + n_3}{n_0} \theta^*_{S3} - \Delta_{S3}\Big)^2 - \Big(\frac{n_1 + n_2 + n_3}{n_0} \theta^*_{S3}\Big)^2}{2z^2},
    \tag{\ref{eq:pse_setup23_criterion2_final}g}
\end{align}
and split the coefficient of $\theta^*_{S3}$ in the first squared term into an integer~(1) and a fractional ($(n_1 + n_2 + n_3)/n_0$) part to arrive at Inequality~\eqref{eq:pse_setup23_criterion2_final}:
\begin{align}
    \frac{2\sigma^2_{C0}}{n_0} >
    \frac{\left(\theta^*_{S3} - \Delta_{S3} + \frac{n_1 + n_2 + n_3}{n_0} \theta^*_{S3}\right)^2 - \left(\frac{n_1 + n_2 + n_3}{n_0} \theta^*_{S3}\right)^2}{2z^2} .
    \nonumber
\end{align}

\section{Dual Control}
We finally clarify the calculations in Section~\ref{sec:expt_comparison_dual_control} of the paper, where we determine the sample size required for Setup 4 (aka a dual control setup) to emerge superior to Setup 3 (a simpler A/B test setup).  In the paper we showed that $\theta^*_{S4} > \theta^*_{S3}$ always holds, and hence Setup 4 will never be superior to Setup 3 under the first evaluation criterion. We thus focus on the second evaluation criterion $\Delta_{S4} - \Delta_{S3} > \theta^*_{S4} - \theta^*_{S3}$, and first show that the criterion is equivalent to Inequality~\eqref{eq:pse_setup34_criterion2_full}. Assuming the ratio of user group sizes ${n_1: n_2: n_3}$ remains unchanged, we then show how
\begin{enumerate}[leftmargin=*]
    \item[(i)] The RHS of the Inequality~\eqref{eq:pse_setup34_criterion2_full} remains a constant and
    \item[(ii)] The LHS of the Inequality~\eqref{eq:pse_setup34_criterion2_full} scales along $O(\sqrt{n})$, where $n$ is the number of users. 
\end{enumerate}
The results mean Setup 4 could emerge superior to Setup 3 if we have sufficiently large number of users. We also show from the inequality that
\begin{enumerate}[leftmargin=*]
    \item[(iii)] The number of users required for a dual control setup to emerge superior to simpler setups is accessible only to the largest organizations and their top affiliates.
\end{enumerate}

\paragraph{The master inequality} We first show the criterion $\Delta_{S4} - \Delta_{S3} > \theta^*_{S4} - \theta^*_{S3}$ is equivalent to
\begin{align}
    &\frac{n_1 \frac{n_2(\mu_{I2}-\mu_{C2}) + n_3(\mu_{I\psi}-\mu_{C3})}{n_2 + n_3} -
           n_2 \frac{n_1(\mu_{I1}-\mu_{C1}) + n_3(\mu_{I\phi}-\mu_{C3})}{n_1 + n_3}}
           {\sqrt{n_1(\sigma^2_{C1} + \sigma^2_{I1}) +
                  n_2(\sigma^2_{C2} + \sigma^2_{I2}) +
                  n_3(\sigma^2_{I\phi} + \sigma^2_{I\psi})}} >
    \tag{\ref{eq:pse_setup34_criterion2_full}}\\
    & \sqrt{2}z \vast[
    \sqrt{
        2 \cdot 
        \frac{
            \begin{array}{l}
                (1 + \frac{n_2}{n_1 + n_3})^2 
                  [n_1(\sigma^2_{C1} + \sigma^2_{I1}) +
                  n_3(\sigma^2_{C3} + \sigma^2_{I\phi})] + \\
                \quad (1 + \frac{n_1}{n_2 + n_3})^2 
                  [n_2(\sigma^2_{C2} + \sigma^2_{I2}) +
                  n_3(\sigma^2_{C3} + \sigma^2_{I\psi})]
            \end{array}}
            {n_1(\sigma^2_{C1} + \sigma^2_{I1}) +
                n_2(\sigma^2_{C2} + \sigma^2_{I2}) +
                n_3(\sigma^2_{I\phi} + \sigma^2_{I\psi})}}
        - 1
    \vast],
    \nonumber
\end{align}
where $z = z_{1-\nicefrac{\alpha}{2}} - z_{1 - \pi_{\textrm{min}}}$. The number of terms involved is large, and hence we first simplify the LHS and RHS independently, and combine them in the final step.

For the LHS (i.e. $\Delta_{S4} - \Delta_{S3}$), we substitute in Equations~\eqref{eq:Delta_S3} and~\eqref{eq:Delta_S4}  to obtain
\begin{align}
    & \frac{n_2(\mu_{I2} \!-\! \mu_{C2}) + n_3(\mu_{I\psi} \!-\! \mu_{C3})}{n_2 + n_3} - 
    \frac{n_1(\mu_{I1} \!-\! \mu_{C1}) + n_3(\mu_{I\phi} \!-\! \mu_{C3})}{n_1 + n_3} - \nonumber\\
    & \frac{n_1(\mu_{C1} - \mu_{I1}) + n_2(\mu_{I2} - \mu_{C2}) + n_3(\mu_{I\psi} - \mu_{I_\phi})}{n_1 + n_2 + n_3}.
    \tag{\ref{eq:pse_setup34_criterion2_full}a}
    \label{eq:pse_setup34_criterion2_fulla}
\end{align}
The expression can be rewritten in terms of multiplicative products between the $n$-terms and the (difference between) $\mu$-terms:
\begin{align}
    & n_1(\mu_{I1} - \mu_{C1})\big[-\textstyle\frac{1}{n_1 + n_3} + \textstyle\frac{1}{n_1 + n_2 + n_3}\big] + \nonumber\\
    & n_2(\mu_{I2} - \mu_{C2})\big[\textstyle\frac{1}{n_2 + n_3} - \textstyle\frac{1}{n_1 + n_2 + n_3}\big] + \nonumber\\
    & n_3\mu_{I\psi}\big[\textstyle\frac{1}{n_2 + n_3} - \textstyle\frac{1}{n_1 + n_2 + n_3}\big] +
    n_3\mu_{I\phi}\big[-\textstyle\frac{1}{n_1 + n_3} + \textstyle\frac{1}{n_1 + n_2 + n_3}\big] + \nonumber\\
    & n_3\mu_{C3}\big[-\textstyle\frac{1}{n_2 + n_3} + \textstyle\frac{1}{n_1 + n_3}\big].
    \tag{\ref{eq:pse_setup34_criterion2_full}b}
    \label{eq:pse_setup34_criterion2_fullb}
\end{align}
We then extract a $1/(n_1 + n_2 + n_3)$ term from Expression~\eqref{eq:pse_setup34_criterion2_fullb}:
\begin{align}
    (n_1 + n_2 + n_3)^{-1}\big[
    & n_1(\mu_{I1} - \mu_{C1})\big(-\textstyle\frac{n_1 + n_2 + n_3}{n_1 + n_3} + 1\big) + \nonumber\\
    & n_2(\mu_{I2} - \mu_{C2})\big(\textstyle\frac{n_1 + n_2 + n_3}{n_2 + n_3} - 1\big)+ \nonumber\\
    & n_3\mu_{I\psi}\big(\textstyle\frac{n_1 + n_2 + n_3}{n_2 + n_3} - 1\big) +
    n_3\mu_{I\phi}\big(-\textstyle\frac{n_1 + n_2 + n_3}{n_1 + n_3} + 1\big) + \nonumber\\
    & n_3\mu_{C3}\big(-\textstyle\frac{n_1 + n_2 + n_3}{n_2 + n_3} + \textstyle\frac{n_1 + n_2 + n_3}{n_1 + n_3}\big)
    \big].
    \tag{\ref{eq:pse_setup34_criterion2_full}c}
    \label{eq:pse_setup34_criterion2_fullc}
\end{align}
This allows us to perform some cancellation with the RHS, which also has a $1/ (n_1 + n_2 + n_3)$ term, in the final step. Noting
\begin{align}
    \frac{n_1 + n_2 + n_3}{n_1 + n_3} = 1 + \frac{n_2}{n_1 + n_3} \;\;\textrm{and}\;\;
    \frac{n_1 + n_2 + n_3}{n_2 + n_3} = 1 + \frac{n_1}{n_2 + n_3},
    \nonumber
\end{align}
we can write the inner square brackets as
\begin{align}
    (n_1 + n_2 + n_3)^{-1}\big[
    & n_1(\mu_{I1} - \mu_{C1})\big(\!-\textstyle\frac{n_2}{n_1 + n_3}\big) +
      n_2(\mu_{I2} - \mu_{C2})\big(\textstyle\frac{n_1}{n_2 + n_3}\big) + \nonumber\\
    & n_3\mu_{I\psi}\big(\textstyle\frac{n_1}{n_2 + n_3}\big) +
      n_3\mu_{I\phi}\big(\!-\textstyle\frac{n_2}{n_1 + n_3}\big) + \nonumber\\
    & n_3\mu_{C3}\big(- 1 - \textstyle\frac{n_1}{n_2 + n_3} + 1 + \frac{n_2}{n_1 + n_3}\big)
    \big],
    \tag{\ref{eq:pse_setup34_criterion2_full}d}
    \label{eq:pse_setup34_criterion2_fulld}
\end{align}
and group the $n_1/(n_2 + n_3)$ and $n_2 / (n_1 + n_3)$ terms to arrive at
\begin{align}
    \frac{1}{n_1 + n_2 + n_3}\Big[
    & \frac{n_1}{n_2 + n_3} \left(n_2(\mu_{I2}-\mu_{C2}) + n_3(\mu_{I\psi}-\mu_{C3})\right) - \nonumber\\
    & \frac{n_2}{n_1 + n_3} \left(n_1(\mu_{I1}-\mu_{C1}) + n_3(\mu_{I\phi}-\mu_{C3})\right) \Big].
    \tag{\ref{eq:pse_setup34_criterion2_full}e}
    \label{eq:pse_setup34_criterion2_fulle}
\end{align}

For the RHS (i.e. $\theta^*_{S4} - \theta^*_{S3}$), we substitute in Equations~\eqref{eq:theta_star_S3} and~\eqref{eq:theta_star_S4} to obtain
\begin{align}
    & 2z\sqrt{
    \frac{n_1(\sigma^2_{C1} \!+\! \sigma^2_{I1}) + n_3(\sigma^2_{C3} \!+\! \sigma^2_{I\phi})}{(n_1 + n_3)^2} +
    \frac{n_2(\sigma^2_{C2} \!+\! \sigma^2_{I2}) + n_3(\sigma^2_{C3} \!+\! \sigma^2_{I\psi})}{(n_2 + n_3)^2}} \nonumber\\
    & - \sqrt{2}z
    \sqrt{\frac{n_1(\sigma^2_{I1} + \sigma^2_{C1}) + n_2(\sigma^2_{C2} + \sigma^2_{I2}) + n_3(\sigma^2_{I\phi} + \sigma^2_{I\psi})}{(n_1 + n_2 + n_3)^2}},
    \tag{\ref{eq:pse_setup34_criterion2_full}f}
    \label{eq:pse_setup34_criterion2_fullf}
\end{align}
where $z = z_{1-\nicefrac{\alpha}{2}} - z_{1 - \pi_{\textrm{min}}}$.
We then extract a $\sqrt{2}z/(n_1 + n_2 + n_3)$ term from Expression~\eqref{eq:pse_setup34_criterion2_fullf} to arrive at
\begin{align}
    \frac{\sqrt{2}z}{n_1 + n_2 + n_3} \Bigg[
    & \sqrt{2} \sqrt{\!\!\!
        \begin{array}{l}
          \big(\frac{n_1 + n_2 + n_3}{n_1 + n_3}\big)^2 [
              n_1(\sigma^2_{C1} \!+\! \sigma^2_{I1}) + n_3(\sigma^2_{C3} \!+\! \sigma^2_{I\phi})] + \\
          \;\; \big(\frac{n_1 + n_2 + n_3}{n_2 + n_3}\big)^2 [
              n_2(\sigma^2_{C2} \!+\! \sigma^2_{I2}) + n_3(\sigma^2_{C3} \!+\! \sigma^2_{I\psi})]
        \end{array}
    } - \nonumber\\
    & \sqrt{n_1(\sigma^2_{C1} + \sigma^2_{I1}) +
            n_2(\sigma^2_{C2} + \sigma^2_{I2}) +
            n_3(\sigma^2_{I\phi} + \sigma^2_{I\psi})}
    \,\Bigg],
    \tag{\ref{eq:pse_setup34_criterion2_full}g}
    \label{eq:pse_setup34_criterion2_fullg}
\end{align}
where $(n_1 + n_2 + n_3)/(n_2 + n_3)$ and $(n_1 + n_2 + n_3)/(n_1 + n_3)$ can also be written as $1 + n_1 / (n_2 + n_3)$ and $1 + n_2 / (n_1 + n_3)$ respectively.

We finally combine both sides of the inequality by taking Expressions~\eqref{eq:pse_setup34_criterion2_fulle} and~\eqref{eq:pse_setup34_criterion2_fullg}:
\begin{align}
    \frac{1}{n_1 + n_2 + n_3}\Big[
    & \frac{n_1}{n_2 + n_3} \left(n_2(\mu_{I2}-\mu_{C2}) + n_3(\mu_{I\psi}-\mu_{C3})\right) - \nonumber\\
    & \;\;\frac{n_2}{n_1 + n_3} \left(n_1(\mu_{I1}-\mu_{C1}) + n_3(\mu_{I\phi}-\mu_{C3})\right) \Big] > \nonumber\\
    \frac{\sqrt{2}z}{n_1 + n_2 + n_3} \Bigg[
    & \sqrt{2} \sqrt{\!\!\!
        \begin{array}{l}
          \big(1 + \frac{n_2}{n_1 + n_3}\big)^2 [
              n_1(\sigma^2_{C1} \!+\! \sigma^2_{I1}) + n_3(\sigma^2_{C3} \!+\! \sigma^2_{I\phi})] + \\
          \quad\big(1 + \frac{n_1}{n_2 + n_3}\big)^2 [
              n_2(\sigma^2_{C2} \!+\! \sigma^2_{I2}) + n_3(\sigma^2_{C3} \!+\! \sigma^2_{I\psi})]
        \end{array}\!\!
    } - \nonumber\\
    & \quad\sqrt{n_1(\sigma^2_{C1} + \sigma^2_{I1}) +
            n_2(\sigma^2_{C2} + \sigma^2_{I2}) +
            n_3(\sigma^2_{I\phi} + \sigma^2_{I\psi})}
    \,\Bigg].
    \tag{\ref{eq:pse_setup34_criterion2_full}h}
    \label{eq:pse_setup34_criterion2_fullh}
\end{align}
Canceling the common $1/(n_1 + n_2 + n_3)$ terms on both sides, and dividing both sides by $\sqrt{n_1(\sigma^2_{C1} \!+\! \sigma^2_{I1}) +
            n_2(\sigma^2_{C2} \!+\! \sigma^2_{I2}) +
            n_3(\sigma^2_{I\phi} \!+\! \sigma^2_{I\psi})}$ 
leads us to Inequality~\eqref{eq:pse_setup34_criterion2_full}:
\begin{align}
    &\frac{n_1 \frac{n_2(\mu_{I2}-\mu_{C2}) + n_3(\mu_{I\psi}-\mu_{C3})}{n_2 + n_3} -
           n_2 \frac{n_1(\mu_{I1}-\mu_{C1}) + n_3(\mu_{I\phi}-\mu_{C3})}{n_1 + n_3}}
           {\sqrt{n_1(\sigma^2_{C1} + \sigma^2_{I1}) +
                  n_2(\sigma^2_{C2} + \sigma^2_{I2}) +
                  n_3(\sigma^2_{I\phi} + \sigma^2_{I\psi})}} >
    \nonumber\\
    & \sqrt{2}z \vast[
    \sqrt{
        2 \cdot 
        \frac{
            \begin{array}{l}
                (1 + \frac{n_2}{n_1 + n_3})^2 
                  [n_1(\sigma^2_{C1} + \sigma^2_{I1}) +
                  n_3(\sigma^2_{C3} + \sigma^2_{I\phi})] + \\
                \quad (1 + \frac{n_1}{n_2 + n_3})^2 
                  [n_2(\sigma^2_{C2} + \sigma^2_{I2}) +
                  n_3(\sigma^2_{C3} + \sigma^2_{I\psi})]
            \end{array}}
            {n_1(\sigma^2_{C1} + \sigma^2_{I1}) +
                n_2(\sigma^2_{C2} + \sigma^2_{I2}) +
                n_3(\sigma^2_{I\phi} + \sigma^2_{I\psi})}}
        - 1
    \vast].
    \nonumber
\end{align}

\paragraph{RHS remains a constant} We then simplify the $\sigma^2$-terms in Inequality~\eqref{eq:pse_setup34_criterion2_full} by assuming that they are similar in magnitude, i.e. 
\begin{align}
    \sigma^2_{C1} \approx \sigma^2_{I1} \approx \cdots \approx \sigma^2_{I\psi} \approx \sigma^2_S \,, \nonumber
\end{align}
and show the RHS of the inequality is equal to
\begin{align}
    \sqrt{2} z \left[ \sqrt{\frac{n_1 + n_2 + n_3}{n_1 + n_3} + \frac{n_1 + n_2 + n_3}{n_2 + n_3}} - 1 \right].
    \tag{\ref{eq:pse_setup34_criterion2_sigmaSimplified}}
\end{align}
As long as the group size ratio $n_1: n_2: n_3$ remains unchanged, Expression \eqref{eq:pse_setup34_criterion2_sigmaSimplified} will remain a constant.
It is safe to apply the simplifying assumption as we know from the evaluation framework specification that there are three classes of parameters in the inequality: the user group sizes ($n$), the mean responses ($\mu$), and the response variances ($\sigma^2$). Among these three classes of parameters, only the user group sizes have the potential to scale in any practical settings, and thus we can effectively treat the means and variances as constants below. 

We begin by substituting $\sigma^2_S$ into Inequality~\eqref{eq:pse_setup34_criterion2_full}:
\begin{align}
    &\frac{n_1 \frac{n_2(\mu_{I2}-\mu_{C2}) + n_3(\mu_{I\psi}-\mu_{C3})}{n_2 + n_3} -
           n_2 \frac{n_1(\mu_{I1}-\mu_{C1}) + n_3(\mu_{I\phi}-\mu_{C3})}{n_1 + n_3}}
           {\sqrt{n_1(2\sigma^2_{S}) +
                  n_2(2\sigma^2_{S}) +
                  n_3(2\sigma^2_{S})}} >
    \nonumber\\
    & \sqrt{2}z \vast[
    \sqrt{
        2 \cdot 
        \frac{
            \begin{array}{l}
                (1 + \frac{n_2}{n_1 + n_3})^2 
                  [n_1(2\sigma^2_{S}) +
                  n_3(2\sigma^2_{S})] + \\
                \quad (1 + \frac{n_1}{n_2 + n_3})^2 
                  [n_2(2\sigma^2_{S}) +
                  n_3(2\sigma^2_{S})]
            \end{array}}
            {n_1(2\sigma^2_{S}) +
                n_2(2\sigma^2_{S}) +
                n_3(2\sigma^2_{S})}}
        - 1
    \vast].
    \tag{\ref{eq:pse_setup34_criterion2_sigmaSimplified}a}
    \label{eq:pse_setup34_criterion2_sigmaSimplified_a}
\end{align}
Moving the common $2\sigma^2_S$ terms out and canceling the common terms in the RHS fraction we have
\begin{align}
    &\frac{n_1 \frac{n_2(\mu_{I2}-\mu_{C2}) + n_3(\mu_{I\psi}-\mu_{C3})}{n_2 + n_3} -
           n_2 \frac{n_1(\mu_{I1}-\mu_{C1}) + n_3(\mu_{I\phi}-\mu_{C3})}{n_1 + n_3}}
           {\sqrt{2\sigma^2_{S}(n_1 + n_2 + n_3)}} >
    \nonumber\\
    & \sqrt{2}z \Bigg[
    \sqrt{
        2 \cdot 
        \frac{ (1 \!+\!\frac{n_2}{n_1 + n_3})^2 (n_1 + n_3) +
               (1 \!+\! \frac{n_1}{n_2 + n_3})^2 (n_2 + n_3)}
            {(n_1 + n_2 + n_3)}}
        - 1
    \Bigg].
    \tag{\ref{eq:pse_setup34_criterion2_sigmaSimplified}b}
    \label{eq:pse_setup34_criterion2_sigmaSimplified_b}
\end{align}
We can already see the LHS of Inequality~\eqref{eq:pse_setup34_criterion2_sigmaSimplified_b} scales along $O(\sqrt{n})$---we will demonstrate this result in greater detail below.

Focusing on the RHS of the inequality, we express the squared terms as rational fractions and divide each term in the numerator by the denominator to obtain
\begin{align}
    \sqrt{2}z \left[\!
    \sqrt{
        2  \left[
        \left(\!\frac{n_1 \!+\! n_2 \!+\! n_3}{n_1 + n_3}\!\right)^{\!2}\!\! \frac{n_1 + n_3}{n_1 \!+\! n_2 \!+\! n_3} \!+\!
        \left(\!\frac{n_1 \!+\! n_2 \!+\! n_3}{n_2 + n_3}\!\right)^{\!2}\!\! \frac{n_2 + n_3}{n_1 \!+\! n_2 \!+\! n_3} \right]}
        \!-\! 1 
    \right].
    \tag{\ref{eq:pse_setup34_criterion2_sigmaSimplified}c}
    \label{eq:pse_setup34_criterion2_sigmaSimplified_c}
\end{align}
Canceling the common $n_1 + n_2 + n_3$ terms leads to that presented in Expression~\eqref{eq:pse_setup34_criterion2_sigmaSimplified}:
\begin{align}
    \sqrt{2} z \left[ \sqrt{\frac{n_1 + n_2 + n_3}{n_1 + n_3} + \frac{n_1 + n_2 + n_3}{n_2 + n_3}} - 1 \right].
    \nonumber
\end{align}

\paragraph{LHS scales along $O(\sqrt{n})$}
We demonstrate the scaling relation between the LHS of Inequality~\eqref{eq:pse_setup34_criterion2_full} and the number of users in each group by simplifying the $n$-terms (but not the $\sigma^2$-terms as above), assuming $n_1 \approx n_2 \approx n_3 \approx n$, and showing the LHS of the inequality is equal to
\begin{align}
    \frac{\sqrt{n} \big((\mu_{I2} - \mu_{C2}) - (\mu_{I1} - \mu_{C1}) + \mu_{I\psi} - \mu_{I\phi}\big)}{2\sqrt{\sigma^2_{C1} + \sigma^2_{I1} + \sigma^2_{C2} + \sigma^2_{I2} + \sigma^2_{I\phi} + \sigma^2_{I\psi}}}.
    \tag{\ref{eq:pse_setup34_criterion2_nSimplified}}
\end{align}
While the relationship (that the LHS of the inequality scales along $O(\sqrt{n})$) is evident by inspecting the LHS of Inequality~\eqref{eq:pse_setup34_criterion2_sigmaSimplified_b} or even Inequality~\eqref{eq:pse_setup34_criterion2_full} itself, we believe the simplification allows us to show the relationship more clearly.

We begin by substituting $n$ into Inequality~\eqref{eq:pse_setup34_criterion2_full} to obtain
\begin{align}
    &\frac{n \frac{n(\mu_{I2}-\mu_{C2}) + n(\mu_{I\psi}-\mu_{C3})}{n + n} -
           n \frac{n(\mu_{I1}-\mu_{C1}) + n(\mu_{I\phi}-\mu_{C3})}{n + n}}
           {\sqrt{n(\sigma^2_{C1} + \sigma^2_{I1}) +
                  n(\sigma^2_{C2} + \sigma^2_{I2}) +
                  n(\sigma^2_{I\phi} + \sigma^2_{I\psi})}} >
    \tag{\ref{eq:pse_setup34_criterion2_nSimplified}a}
    \label{eq:pse_setup34_criterion2_nSimplified_a}\\
    & \sqrt{2}z \vast[
    \sqrt{
        2 \cdot 
        \frac{
            \begin{array}{l}
                (1 + \frac{n}{n + n})^2 
                  [n(\sigma^2_{C1} + \sigma^2_{I1}) +
                  n(\sigma^2_{C3} + \sigma^2_{I\phi})] + \\
                \quad (1 + \frac{n}{n + n})^2 
                  [n(\sigma^2_{C2} + \sigma^2_{I2}) +
                  n(\sigma^2_{C3} + \sigma^2_{I\psi})]
            \end{array}}
            {n(\sigma^2_{C1} + \sigma^2_{I1}) +
                n(\sigma^2_{C2} + \sigma^2_{I2}) +
                n(\sigma^2_{I\phi} + \sigma^2_{I\psi})}}
        - 1
    \vast].
    \nonumber
\end{align}
Moving the common $n$-terms out and canceling them in the fractions where appropriate lead to
\begin{align}
    &\frac{\sqrt{n} \,\frac{1}{2}\left[((\mu_{I2}-\mu_{C2}) + (\mu_{I\psi}-\mu_{C3})) -
           ((\mu_{I1}-\mu_{C1}) + (\mu_{I\phi}-\mu_{C3})) \right]}
           {\sqrt{\sigma^2_{C1} + \sigma^2_{I1} +\sigma^2_{C2} + \sigma^2_{I2} +\sigma^2_{I\phi} + \sigma^2_{I\psi}}} >
    \nonumber\\
    & \sqrt{2}z \vast[
    \sqrt{
        2 \cdot 
        \frac{
            \begin{array}{l}
                (1 + \frac{1}{2})^2 
                  (\sigma^2_{C1} + \sigma^2_{I1} +
                   \sigma^2_{C3} + \sigma^2_{I\phi}) + \\
                \quad (1 + \frac{1}{2})^2 
                  (\sigma^2_{C2} + \sigma^2_{I2} +
                   \sigma^2_{C3} + \sigma^2_{I\psi})
            \end{array}}
            {\sigma^2_{C1} + \sigma^2_{I1} +
                \sigma^2_{C2} + \sigma^2_{I2} +
                \sigma^2_{I\phi} + \sigma^2_{I\psi}}}
        - 1
    \vast],
    \tag{\ref{eq:pse_setup34_criterion2_nSimplified}b}
    \label{eq:pse_setup34_criterion2_nSimplified_b}
\end{align}
where the LHS is equal to Expression~\eqref{eq:pse_setup34_criterion2_nSimplified} as claimed above.

It is clear that there are no $n$-terms left on the RHS of Inequality~\eqref{eq:pse_setup34_criterion2_nSimplified}, and hence the RHS remains a constant as shown previously. Setting up the inequality to demonstrate the third result---that the number of users required for a dual control setup to emerge superior is large---we further simplify the RHS of the inequality by rearranging the terms in the square root:
\begin{align}
    &\frac{\sqrt{n} \left[((\mu_{I2}-\mu_{C2}) + (\mu_{I\psi}-\mu_{C3})) -
           ((\mu_{I1}-\mu_{C1}) + (\mu_{I\phi}-\mu_{C3})) \right]}
           {2\sqrt{\sigma^2_{C1} + \sigma^2_{I1} +\sigma^2_{C2} + \sigma^2_{I2} +\sigma^2_{I\phi} + \sigma^2_{I\psi}}} >
    \nonumber\\
    & \sqrt{2}z \Bigg[
    \sqrt{
        2 \bigg(\frac{3}{2}\bigg)^2
        \bigg(1 + \frac{2\sigma^2_{C3}}
                  {\sigma^2_{C1} + \sigma^2_{I1} + \sigma^2_{C2} +
                   \sigma^2_{I2} +\sigma^2_{I\phi} + \sigma^2_{I\psi}}\bigg)}
        - 1
    \Bigg].
    \tag{\ref{eq:pse_setup34_criterion2_nSimplified}c}
    \label{eq:pse_setup34_criterion2_nSimplified_c}
\end{align}

\paragraph{Required number of users is large}
We finally show that while Setup 4 could emerge superior to Setup 3 as the number of users increase, the number of users required is high. We do so by assuming both the $\sigma^2$- and $n$-terms are similar in magnitude, i.e. $\sigma^2_{C1} \approx \sigma^2_{I1} \approx \cdots \approx \sigma^2_{I\psi} \approx \sigma^2_S$ and $n_1 \approx n_2 \approx n_3 \approx n$, and show that Inequality~\eqref{eq:pse_setup34_criterion2_full} is equivalent to
\begin{align}
    n > \left(2\sqrt{12}\left(\sqrt{6} - 1\right)z\right)^2 \frac{\sigma^2_S}{\Delta^2} \,,
    \tag{\ref{eq:pse_setup34_criterion2_bothSimplified}}
\end{align}
where $\Delta = (\mu_{I2} - \mu_{C2}) - (\mu_{I1} - \mu_{C1}) + \mu_{I\psi} - \mu_{I\phi}$ is the actual effect size difference between Setups 4 and 3. Note we are determining when Setup 4 is superior to Setup 3 under the second evaluation criterion---that \emph{the gain in actual effect} is greater than the loss in sensitivity---and thus assume $\Delta$ is positive.

The equivalence can be shown by substituting $\sigma^2_S$ into Inequality~\eqref{eq:pse_setup34_criterion2_nSimplified_c}, which already assumes the $n$-terms are similar in magnitude:\footnote{Alternatively we can substitute $n$ into Inequality~\eqref{eq:pse_setup34_criterion2_sigmaSimplified_b}, which already assumes the $\sigma^2$-terms are similar in magnitude. Simplifying the resultant inequality would yield the same end result.}
\begin{align}
    &\frac{\sqrt{n} \left[((\mu_{I2}-\mu_{C2}) + (\mu_{I\psi}-\mu_{C3})) -
           ((\mu_{I1}-\mu_{C1}) + (\mu_{I\phi}-\mu_{C3})) \right]}
           {2\sqrt{6\sigma^2_S}} >
    \nonumber\\
    & \sqrt{2}z \Bigg[
    \sqrt{
        2 \bigg(\frac{3}{2}\bigg)^2
        \bigg(1 + \frac{2\sigma^2_{S}}
                  {6\sigma^2_{S}}\bigg)}
        - 1
    \Bigg].
    \tag{\ref{eq:pse_setup34_criterion2_bothSimplified}a}
    \label{eq:pse_setup34_criterion2_bothSimplified_a}
\end{align}
Noting the expression within the LHS square bracket is equal to $\Delta$, we simplify the expression within the RHS square root, and move every non-$n$ term to the RHS of the inequality to obtain
\begin{align}
    \sqrt{n} > \sqrt{2}{z}\left[\sqrt{6}-1\right]\frac{2\sqrt{6\sigma^2_S}}{\Delta}.
    \tag{\ref{eq:pse_setup34_criterion2_bothSimplified}b}
    \label{eq:pse_setup34_criterion2_bothSimplified_b}
\end{align}
As all quantities in the inequality are positive, we can square both sides and consolidate the coefficients on the RHS to arrive at Inequality~\eqref{eq:pse_setup34_criterion2_bothSimplified}:
\begin{align}
    n > \left(2\sqrt{12}\left(\sqrt{6} - 1\right)z\right)^2 \frac{\sigma^2_S}{\Delta^2}.
    \nonumber
\end{align}

\end{document}